%% file: arXiv-Resubmission_OptimalSmearing.tex
\documentclass[twocolumn,aps,pra,groupedaddress,superscriptaddress,amsmath,amssymb,noeprint,nofootinbib,floatfix,10pt]{revtex4-2}
\input{preamble_packages}
\input{preamble_commands}

\makeatletter 

\renewcommand\onecolumngrid{% <<<<<<
\do@columngrid{one}{\@ne}%
\def\set@footnotewidth{\onecolumngrid}% <<<<<<<<<<<<<<<<
\def\footnoterule{\kern-6pt\hrule width 1.5in\kern6pt}%
}

\renewcommand\twocolumngrid{% <<<<<<
\def\footnoterule{% restore rule
\dimen@\skip\footins\divide\dimen@\thr@@
\kern-\dimen@\hrule width.5in\kern\dimen@}
\do@columngrid{mlt}{\tw@}
}%

\makeatother    
\begin{document}

\title{Diffusion Minimization via Optimal Smearing in Collapse and Hybrid Classical-Quantum Gravitational Models}

\author{Nicol\`{o} Piccione}
\email{nicolo.piccione@units.it}
\affiliation{Department of Physics, University of Trieste, Strada Costiera 11, 34151 Trieste, Italy}
\affiliation{Istituto Nazionale di Fisica Nucleare, Trieste Section, Via Valerio 2, 34127 Trieste, Italy}

\begin{abstract}
Spontaneous diffusion (i.e., non-conservation of energy) is a prominent, testable prediction of collapse and hybrid classical-quantum gravitational models. Without smearing of the mass density operator, the associated heating (or energy increase) rate diverges, yet the smearing distribution is arbitrary and, on scales much larger than the smearing length $r_C$, much of the phenomenology is expected to be insensitive to this choice.
We propose to resolve this arbitrariness as follows: for a fixed $r_C$, select the distribution that minimizes the heating rate. Conceptually, this should identify the minimal deviation from standard quantum mechanics and provide models that, once experimentally refuted, would strongly disfavor all variants with different distributions. We apply this approach to the most investigated collapse models: GRW (for Ghirardi-Rimini-Weber), CSL (for Continuous Spontaneous Localization), and DP (for Diósi-Penrose).  Notably, the Gaussian is optimal only for the GRW case. Finally, we apply it to the Tilloy-Diósi hybrid classical-quantum model of Newtonian gravity, leading to the minimally deviating variant of it. This version of the model is entirely determined by only one free parameter $r_C$ and, if experimentally refuted, would strongly disfavor any other version of it.
\end{abstract}

\maketitle

\section{Introduction\label{Sec:Introduction}}

Two highly relevant problems that affect quantum mechanics are the measurement problem~\cite{Book_Bell2004Speakable,Book_Norsen2017foundations,Book_Durr2020understanding,Book_Tumulka2022Foundations} and finding a way to merge quantum mechanics with gravity~\cite{Kiefer2023Review_QuantumGravityUnfinished}. A possible solution to the former is provided by spontaneous collapse models~\cite{Review_Bassi2003Dynamical,Review_Bassi2013Models,Book_Tumulka2022Foundations,Review_Bassi2023CollapseModels}, which challenge the universal validity of the superposition principle. 
%These models can be seen either as phenomenological ones stemming from more fundamental physical theories~\cite{Review_Bassi2013Models} or as well-defined physical theories on their own~\cite{Book_Tumulka2022Foundations}.
One possible approach to the second problem, historically pursued by a minority of physicists but recently gaining momentum, consists of keeping the gravitational field (or curved spacetime) classical and understanding how such a classical system may interact with quantum matter~\cite{Kafri2014LOCCGravity,Kafri2015BoundsNewtonianGravity,Tilloy2016CSLGravity,Tilloy2017LeastDecoherence,GaonaReyes2021GravitationalFeedback,Oppenheim2023PostQuantum,Piccione2025NewtonianPSL}. In fact, the very (quantum or classical) nature of gravity is still unclear and experiments have been proposed~\cite{Bose2017GravityExperiment,Marletto2017GravityExperiment,Lami2024GravityTesting,Angeli2025ProbingQuantumNatureGravity,Review_Bose2025GravityExperimentalTests} to determine it.

The dynamics of both collapse and hybrid models is, necessarily, nonunitary and stochastic, entailing energy non-conservation. Collapse models aim to substitute the measurement postulate of standard quantum mechanics by modifying the dynamical law itself~\cite{Review_Bassi2003Dynamical,Review_Bassi2013Models}. Since the measurement dynamics is stochastic and nonlinear, the spontaneous collapse dynamics is stochastic and nonunitary~\cite{Review_Bassi2003Dynamical,Review_Bassi2013Models}. This modification is negligible for small quantum systems but becomes prominent for increasingly large systems~\cite{Review_Bassi2003Dynamical,Review_Bassi2013Models}, so that macroscopic objects are basically never allowed to be in a macroscopic superposition of positions. For collapse models to work as intended, they must be diffusive~\cite{Donadi2023DiffusiveCollapse}, that is, characterized by a positive energy increase rate, or heating rate. This lack of energy conservation constitutes one of the main theoretically observable differences from standard quantum mechanics. This requirement plausibly extends to hybrid classical-quantum models as well, because they are believed to be inherently irreversible~\cite{Galley2023ConsistentQuantumClassicalGravity,Oppenheim2023PostQuantum} and, in particular, to include spontaneous collapse in their dynamics~\cite{Tilloy2024HybridDynamics,Barchielli2024HybridDynamics}.
Roughly speaking, if the classical gravitational field depends on where masses are, it also reveals where they are, thus inducing a spontaneous spatial localization of the wavefunction.

In order to make the spontaneous heating not divergent, it is generally necessary to perform some kind of smearing operation\footnote{To understand why this is so, consider the following simple example: a particle perfectly localizes so that $\Delta x \rightarrow 0$, where $\Delta x$ is its spatial dispersion. Then, by Heisenberg's uncertainty principle, it would have infinite energy immediately after the spontaneous collapse.}. Usually, this is done by smearing the matter density operator by means of a Gaussian of length $r_C$. The choice of using a Gaussian is largely conventional; for any reasonable smearing, large-distance features are expected to be insensitive to the smearing distribution profile. In particular, one expects the dynamics to become non-entangling~\cite{Trillo2025DiosiPenroseEntanglement,Angeli2025EntanglementHybridGravity}, long-range spatial decoherence\footnote{In this paper, to comply with the general linguistic usage of different communities,
we will use the term ``collapse rate'' when dealing with collapse models and ``decoherence rate'' when dealing with hybrid gravitational models.} to saturate at the no-smearing limit~\cite{Toros2018BoundsCalculations,Piccione2025ExploringMassDependence,Figurato2024DPEffectiveness,Piccione2025NewtonianPSL}, and, in hybrid models, the gravitational interaction to reduce to Newtonian gravity~\cite{Tilloy2016CSLGravity,Piccione2025NewtonianPSL}. With the typical values chosen for $r_C$ (i.e., $r_C \sim 10^{-7} \textrm{m}$), the spontaneous heating predicted by collapse and hybrid models is compatible with current experimental bounds because it is typically tiny, making finite-time energy growth difficult to observe~\cite{Carlesso2022Present,Piccione2025ExploringMassDependence}. However, it may well be that, for a given model and a value of $r_C$, a lower heating rate is attained by employing a different smearing distribution than the Gaussian.

\input{table}

We propose a simple way to fix the smearing arbitrariness: given a spatial variance $r_C^2$ of the smearing distribution, one should choose the profile that minimizes the heating rate. This allows us to set up an optimization problem for all collapse and hybrid models where, mathematically, it is like the (smeared) mass density operator is continuously and weakly monitored. We find the explicit and unique solutions to the optimization problem for the Ghirardi-Rimini-Weber (GRW)~\cite{Ghirardi1986Unified}, the Continuous Spontaneous Localization (CSL)~\cite{Pearle1989CSL,Ghirardi1990_CSL,Pearle1994MassCSL}, and the Diósi-Penrose (DP)~\cite{Diosi1987Universal,Diosi1989Models,Penrose1996gravity,Penrose2014Gravitization} spontaneous collapse models; the Gaussian distribution results optimal only in the GRW case.
Regarding hybrid gravitational models, Ref.~\cite{Tilloy2017LeastDecoherence} proposed a ``Principle of Least Decoherence'' (PLD) which, under certain assumptions, identifies a precise hybrid gravitational model, up to choice of the smearing distribution, which we denote as the Tilloy-Diósi (TD) model. For this model, the optimal smearing distribution coincides with that found for the DP model, thus providing a model entirely characterized by a single parameter $r_C$. All possible values of this parameter can be experimentally falsified: lower bounds come from compatibility of heating rate and related phenomena with experimental data~\cite{Donadi2021UndergroundTest} while upper bounds can be put by observing no deviations from Newton's law of gravitation at smaller and smaller length-scales. We also discuss what happens if the assumption of equal smearing for the spontaneous measurement and gravitational feedback dynamics does not hold.

The paper is organized as follows.
In Sec.~\ref{Sec:CollapseModels}, we show how to map the goal of minimizing the diffusion rate for a given collapse radius into a minimization problem. We do this for the GRW, CSL, and DP collapse models and then solve the optimization problem, thus finding the optimal smearing profiles.
In Sec.~\ref{Sec:HybridModels}, we do the same for the Tilloy-Diósi model of hybrid Newtonian gravity~\cite{Tilloy2016CSLGravity,Tilloy2017LeastDecoherence}.
In Sec.~\ref{Sec:Generalization}, we show how the same minimization criterion can be formulated for the broader class of collapse and hybrid models based on continuous weak monitoring of the mass density.
Then, in Sec.~\ref{Sec:ConnectionExperimentalBounds}, we discuss the connection of our findings with experimental bounds.
In Sec.~\ref{Sec:Discussion}, we discuss the role of the variance constraint, the sense in which large-scale phenomenology is insensitive to the smearing profile, and the relation between the GRW and CSL optimal smearing profiles.
Finally, Sec.~\ref{Sec:Conclusions} summarizes the conclusions.

\section{Collapse Models\label{Sec:CollapseModels}}

Consider a generic system of $N$ particles governed by the Hamiltonian
\begin{equation}\label{eq:StandardHamiltonian}
\hH = \sum_{j=1}^N \frac{\hbp_j^2}{2 m_j} + V(\hbq_1,\dots,\hbq_N),
\end{equation}
where $m_j$ is the $j$-th particle's mass, $\hbp_j$ its momentum operator, and $\hbq_j$ its position operator. The master equation of the considered collapse models is generically given by
\begin{equation}
\dot{\rho}_t = -\frac{i}{\hbar}\comm{\hH}{\rho_t} + \mcL_C [\rho_t],
\end{equation}
where the actual form of $\mcL_C [\rho_t]$ for each model can be seen in Table~\ref{tab:my-table}.

The quantity we want to minimize is $\dot{E}_t := \dv{t} \langle \hH\rangle_t $, which, since the potential commutes with all operators in $\mcL_C [\rho_t]$, can be computed (in all three cases) as a sum of single particle contributions~\cite{Review_Bassi2003Dynamical,Review_Bassi2013Models}. Notably, $\dot{E}_t$ is a state-independent quantity proportional to a functional of the smearing distribution $g_{r_C}$ (see second row of Table~\ref{tab:my-table}). The constraints under which we minimize $\dot{E}_t$ are
\begin{equation}\label{eq:SmearingConstraints}
g_{r_C} (\bx) \geq 0, 
\
\int \dd[3]{\bx} g_{r_C} (\bx)  = 1,
\
\int \dd[3]{\bx} \bx^2 g_{r_C} (\bx)  = 3 r_C^2.
\end{equation}
Notice that, in stating the variance constraint, we are implicitly assuming that $\int \dd[3]{\bx} g_{r_C} (\bx) \bx = 0$. Indeed, it makes physical sense to assume that the smearing is centered.

To minimize the energy increase in the GRW model it is sufficient to notice that (see more details in Appendix~\ref{APPSec:GRW_Appendix})
\begin{equation}
\dot{E}_t
\propto
I[f]
= \frac{1}{2}\int \dd[3]{\bk} \bk^2 \abs{\tl{f}(\bk)}^2,	
\end{equation}
where $\tl{f}(\bk) := (2\pi)^{-3/2}\int \dd[3]{\bx} f(\bx) e^{-i \bk \cdot \bx}$. Thus, by making the identification $\psi (\bx) = \sqrt{g_{r_C} (\bx)}$, what one searches is the real, non-negative, wavefunction $\psi (\bx)$ such that the product of the variances of $\abs{\psi (\bx)}^2$ and $\abs{\tl{\psi} (\bk)}^2$ is minimized. This is an already solved problem as it reduces to that of saturating the three-dimensional uncertainty principle~\cite{Folland1997SurveyUncertaintyPrinciple,McCurdy2021StabilityHPWInequality,Fathi2021ShortProofHPWInequality,Review_Dodonov2025UncertaintyRelations}. The solution is given by Gaussian wavefunctions~\cite{McCurdy2021StabilityHPWInequality,Fathi2021ShortProofHPWInequality}. Therefore, $g_{r_C} (\bx)$ is a centered Gaussian distribution of variance $r_C^2$.

For the CSL model (more details in Appendix~\ref{APPSec:CSL_Appendix}), one can first make use of the Pólya–Szegő rearrangement inequality to show that the optimal distribution is radial and decreasing. Thus, it is a distribution which, in spherical coordinates, either has infinite support or its support goes from $0$ to $R>0$. We include both cases by considering also $R=+\infty$. As $\dot{E}_t \propto I[g_{r_C}]$, we only need to minimize $I[g_{r_C}]$, which in spherical coordinates reads
\begin{equation}
I[g_{r_C}] = 2\pi \int_0^R \dd{r} r^2 [g'(r)]^2.
\end{equation}
Moreover, the constraints can be written as
\begin{equations}
4\pi \int_0^R \dd{r} r^2 g_{r_C}(r) &=1,
\\
4\pi \int_0^R \dd{r} r^4 g_{r_C}(r) &=3 r_C^2,
\\
g_{r_C}(r) &\geq 0,
\end{equations}
and the Lagrange multiplier technique gives
\begin{multline}
\mcL[g_{r_C}] = - \lambda - 3\mu r_C^2 + 4\pi 
\times \\ \times
\int_0^{R} \prt{\frac{1}{2}r^2 [\partial_r g_{r_C} (r)]^2 + \lambda r^2 g_{r_C}(r) + \mu r^4 g_{r_C} (r)}\dd{r},
\end{multline}
where $\lambda$ and $\mu$ are, here, the Lagrange multipliers. Finding the minimum and exploiting the constraints leads to
\begin{equation}
g_{r_C}(\bx) = \frac{105}{32 \pi R^7}\prt{R^2-\bx^2}^2 \Theta(R - \abs{\bx})\vert_{R=3 r_C}.
\end{equation}

The solution method for the DP case is similar to the one used for CSL (see Appendix~\ref{APPSec:DP_Appendix} for details). In particular, the functional simplifies to
\begin{equation}
I_{\rm DP}[g_{r_C}] = \pi\int \dd[3]{\bx} g_{r_C}^2 (\bx).    
\end{equation}
The optimal distribution has to be, again, radial and decreasing. The Lagrange multiplier technique leads to 
\begin{multline}
\mcL[g_{r_C}] = - \lambda - 3 \mu r_C^2 +
\\
+4\pi\int_0^{R} \prt{\pi r^2 [g_{r_C} (r)]^2 + \lambda r^2 g_{r_C}(r) + \mu r^4 g_{r_C} (r)}\dd{r}.
\end{multline}
The fixed-variance solution is
\begin{equation}
g_{r_C} (\bx) 
= 
\frac{15}{8 \pi R^5}\prt{R^2 - \bx^2}\Theta(R - \abs{\bx})\vert_{R=\sqrt{7}r_C}.
\end{equation}

Notably, on physical grounds, one could have also added spherical symmetry to the list of constraints. However, this is not necessary; the optimal distributions have to be radial and decreasing. The results of our calculations are in the third row of Table~\ref{tab:my-table}. Moreover, in the last row we show how much higher the heating rate is when using the Gaussian distribution in place of the optimal one.
Notably, and perhaps surprisingly, these deviations are of order unity for the DP and CSL models.

\section{Hybrid Gravitational Models\label{Sec:HybridModels}}

In Ref.~\cite{Tilloy2016CSLGravity}, Tilloy and Diósi proposed a general prescription to introduce classical gravity starting from a spontaneous collapse model in which the (smeared) mass density is weakly and continuously monitored. The prescription consists of treating gravity as a feedback mechanism based on the mass density readings due to the spontaneous collapse dynamics\footnote{This approach based on measurement and feedback has proven to be equivalent to other ones~\cite{Tilloy2024HybridDynamics} like that of Ref.~\cite{Oppenheim2023PostQuantum}.}. It is usually assumed that the smearing distributions used for the measurement and feedback dynamics are equal, but this is not necessary, in principle. Therefore, we do not make this assumption. For convenience, we summarize the approach of Ref.~\cite{Tilloy2016CSLGravity} in Appendix~\ref{APPSec:WeaklyContinuousMonitoring}, and report here the effective master equation of the quantum system:
\begin{equation}\label{eq:HybridModelsEquation}
\dot{\rho}_t = -\frac{i}{\hbar}\comm{\hH+\hV_{r_C,r_G}}{\rho_t} + \mcL_C [\rho_t] + \mcL_G [\rho_t],
\end{equation}
where
\begin{equations}\label{eq:HybridModelsEquation_Components}
\hV_{r_C,r_G} 
:&= \int \dd[3]{\bx}\dd[3]{\by} \prt{\frac{-G}{\abs{\bx-\by}}} \frac{\hmu_{r_C} (\bx)\hmu_{r_G}(\by)}{2},
\\
\mcL_C [\rho_t] &= - \frac{1}{2} \int \dd[3]{\bx}\dd[3]{\by} \gamma_C (\bx,\by) \comm{\hmu_{r_C} (\bx)}{\comm{\hmu_{r_C} (\by)}{\rho_t}},
\\
\mcL_G [\rho_t] &= - \frac{1}{2} \int \dd[3]{\bx}\dd[3]{\by} \gamma_G (\bx,\by) \comm{\hmu_{r_G} (\bx)}{\comm{\hmu_{r_G} (\by)}{\rho_t}}.
\end{equations}
In the above equations, $\hV_{r_C,r_G}$ is the gravitational Hamiltonian and both $\gamma_C (\bx,\by)$ and $\gamma_G (\bx,\by)$ are translation invariant correlators, whose connection is more easily seen in Fourier space (see Appendix~\ref{APPSec:WeaklyContinuousMonitoring} for more details):
\begin{equation}
\tl{\gamma}_G (\bk) = \frac{G^2}{2\pi \hbar^2 \bk^4} \frac{1}{\tl{\gamma}_C (\bk)},
\end{equation}
where $\tl{f}(\bk) = (2\pi)^{-3/2}\int \dd[3]{\bx} f(\bx) e^{-i \bk \cdot \bx}$ and $G$ is Newton's constant.
In Eqs.~\eqref{eq:HybridModelsEquation} and~\eqref{eq:HybridModelsEquation_Components}, $\hmu_{r_C} (\bx)$ is the same kind of smeared mass density operator\footnote{When dealing with a fixed number $N$ of particles, the smeared mass density operator with smearing distribution $g_\sigma (\bx)$ reads $\hmu_\sigma = \sum_j m_j g_\sigma (\bx-\hbq_j)$, where $m_j$ is the mass of the $j$-th particle and $\hbq_j$ its position operator.} that appears in the CSL or DP models, while $\hmu_{r_G} (\bx)$ is the smeared mass density operator with smearing distribution $g_{r_G} (\bx)$, in general different from $g_{r_C} (\bx)$. As for collapse models, the smearing $g_{r_G}$ entering both the gravitational Hamiltonian and the new noise term may be necessary to avoid divergences of the heating rate~\cite{Tilloy2016CSLGravity,GaonaReyes2021GravitationalFeedback}. Of course, $\hV_{r_C,r_G}$ reduces to the standard Newtonian interaction Hamiltonian when considering distances much higher than $r_C$ and $r_G$.
Notice that setting $\hV_{r_C,r_G} $ and $\mcL_G [\rho_t]$ to zero, Eq.~\eqref{eq:HybridModelsEquation} reduces to the master equation of a spontaneous collapse model. In particular, CSL is recovered by setting $\gamma_C (\bx,\by) = m_0^{-2}\gamma_{\rm CSL}\delta (\bx-\by)$, and DP by setting instead $\gamma_{C} (\bx,\by) = (G/2\hbar)\abs{\bx-\by}^{-1}$. 

Assuming that the smearing for the spontaneous measurement part of the dynamics and the gravitational feedback part is the same, Tilloy and Diósi proposed to single out a correlator $\gamma_C (\bx,\by)$ by following the
``Principle of Least Decoherence'' (PLD)~\cite{Tilloy2017LeastDecoherence}. 
This leads, up to the choice of smearing distribution, to the model that we call Tilloy-Diósi (TD) model, where one has that $\mcL_G [\rho_t] = \mcL_C [\rho_t]$, with $\mcL_C [\rho_t]$ being that of the DP model (see Table~\ref{tab:my-table}). 
In fact, already for a single particle, one can see that minimizing the decoherence rate in Fourier space gives (see Appendix~\ref{APPSec:WeaklyContinuousMonitoring})
\begin{equation}\label{eq:PLD_MinimizationCorrelator}
\tl{\gamma}_C (\bk) = \frac{G}{\hbar \sqrt{2\pi} \bk^2} \frac{\abs{\tl{g}_{r_G} (\bk)}}{\abs{\tl{g}_{r_C} (\bk)}},
\quad
\tl{\gamma}_G (\bk) = \frac{G}{\hbar \sqrt{2\pi} \bk^2} \frac{\abs{\tl{g}_{r_C} (\bk)}}{\abs{\tl{g}_{r_G} (\bk)}},
\end{equation}
which returns the DP model correlator for both $\gamma_C$ and $\gamma_G$ when $g_{r_G}=g_{r_C}$. Thus, the heating rate is calculated exactly as for the DP model and the minimizing distribution is also the same (see Table~\ref{tab:my-table}). Our proposal, in addition to the PLD, singles out a specific variant of the TD model by singling out the optimal smearing distribution associated to the DP noise correlator.
This yields a hybrid gravity model entirely determined by a single parameter $r_C$. Importantly, both upper and lower bounds can be put on this parameter~\cite{Tilloy2016CSLGravity,Tilloy2017LeastDecoherence}, thus making it possible to completely exclude it experimentally. Since our additional demand leads to what we deem to be the minimal deviation from standard quantum mechanics, we argue that the exclusion of the TD model with the optimal smearing of Table~\ref{tab:my-table} would strongly disfavor the TD model for any other smearing profile.

We can also drop the assumption that $g_{r_G} = g_{r_C}$. In this case, Eq.~\eqref{eq:PLD_MinimizationCorrelator} shows that the PLD no longer singles out the DP model. Then, with a general noise correlation, the optimization problem can be solved separately for the spontaneous measurement part and for the gravitational feedback part (see next section). If the noise correlation is chosen to be the DP one, the two optimization problems will be functionally equal and will give the same solution of Table~\ref{tab:my-table}, with $r_C$ and $r_G$ quantifying the distribution variances, respectively.

\section{Generalization\label{Sec:Generalization}}

We found the optimal distributions for the GRW, CSL, DP, and TD models. However, our approach can be applied to any model described by Eq.~\eqref{eq:HybridModelsEquation}. 
In fact, for any such model, one has that $\dot{E}_t = \dot{E}^{(C)}_t+\dot{E}^{(G)}_t$, where ($A$ denoting $C$ or $G$) $\dot{E}^{(A)}_t = \hbar^2 M I_{\gamma_{A}} [g_{r_{A}}]$, with 
\begin{equation}\label{eq:GenericHeatingRateFunctional}
I_{\gamma_{A}} [g_{r_{A}}]
=
\frac{1}{2}\int \dd[3]{\bx}\dd[3]{\by} \gamma_{A} (\bx,\by) \prtq{\nabla g_{r_{A}}(\bx)}\cdot\prtq{\nabla g_{r_{A}}(\by)}.
\end{equation}
So, for any model (collapse or hybrid) described by Eqs.~\eqref{eq:HybridModelsEquation} and~\eqref{eq:HybridModelsEquation_Components} one can set up a problem with state-independent functionals to minimize, leading to a state-independent optimal smearing distribution $g_{r_A}^{(\gamma_A)} (\bx)$.
Therefore, our proposal provides a universal way to associate with any correlator $\gamma_A$ a smearing distribution that minimizes the spontaneous diffusion.
Models such as the Poissonian Spontaneous Localization model~\cite{Piccione2023Collapse,Piccione2025ExploringMassDependence} and its gravitational version~\cite{Piccione2023Collapse,Piccione2025NewtonianPSL} do not have, in general, a state-independent heating rate. We explore how to adapt this paper's proposal to their case in Ref.~\cite{Piccione2026EnergyIncreaseGPSL}.

\section{Connection with experimental bounds\label{Sec:ConnectionExperimentalBounds}}

Collapse and hybrid models predict deviations from standard quantum mechanics which can be used to set bounds on their parameters. Among these predictions, the spontaneous emission of radiation~\cite{Fu1997SpontaneousRadiation,Donadi2014RadiationEmission,Donadi2015Radiation} and the diffusion of a rigid body's center of mass~\cite{Carlesso2016ExperimentalBounds} lead to the most stringent bounds~\cite{Carlesso2022Present,MAJORANACollaboration2022WaveFunctionCollapse,XENONCollaboration2026CollapseConstraints,Piccione2025ExploringMassDependence}. The spontaneous increase of the \emph{total} energy of a system has also been used to place such bounds~\cite{Tilloy2019NeutronStarHeating,Adler2019NeutronStarSpontaneousHeating,Piccione2026EnergyIncreaseGPSL}, despite the fact that (at least for CSL and DP) it does not lead to the most stringent ones.

The bounds on CSL, DP, and TD are obtained by comparing experimental and observational data with theoretical predictions, under the assumption that the smearing distribution is a Gaussian. Thus, strictly speaking, they lack generality. To make them more general, the method used in this paper could be applied. That is, one should first derive the theoretical deviation from standard quantum mechanics with a generic smearing and only at the end search for the smearing profile (at fixed smearing length) that minimizes that deviation. This would make the bounds more general.

In Appendix~\ref{APPSec:ConnectionExperimentalBounds}, we show that, in relevant cases, minimizing the spontaneous diffusion also minimizes the emission of radiation due to the collapse and hybrid dynamics.
In particular, for a single particle one gets that (cf. the general model of Sec.~\ref{Sec:Generalization})
\begin{equation}
\dv{\Gamma (t)}{\omega}
=
\frac{1}{\omega}\times
\frac{q^2 \hbar}{3 \pi^2 \varepsilon_0 c^3} \prt{I_{\gamma_C}[g_{r_C}] + I_{\gamma_G}[g_{r_G}]},
\end{equation}
where $\dd{\Gamma (t)}/\dd{\omega}$ is the emission rate at a given frequency $\omega$, $q$ is the particle's charge, $c$ is the speed of light, and $\varepsilon_0$ is the vacuum permittivity.
Following the lines of Ref.~\cite{Donadi2021NovelCSLBounds} (see Appendix~\ref{APPSec:ConnectionExperimentalBounds} for more details) one can show that under reasonable approximations $\dd{\Gamma (t)}/\dd{\omega} \propto I_{\gamma_C}[g_{r_C}] + I_{\gamma_G}[g_{r_G}]$ even for a large rigid body.
Since spontaneous radiation emission constitutes the phenomenon that allows to put the strongest bounds on the DP and TD models~\cite{MAJORANACollaboration2022WaveFunctionCollapse,XENONCollaboration2026CollapseConstraints}, its optimization equivalence with minimization of diffusion strengthens our argument that the models we find constitute the minimally deviating variant of such models.

In Appendix~\ref{APPSec:ConnectionExperimentalBounds}, we also compute the center-of-mass diffusion of a large macroscopic body for the general model of Sec.~\ref{Sec:Generalization}. The result is that the diffusion rate separates into two pieces:
\begin{equation}
\dot{E}^{\rm CoM}_t
\simeq
\dot{E}^{\rm CoM}_t\vert_{V} + \dot{E}^{\rm CoM}_t\vert_{\partial V},
\end{equation}
where $V$ denotes the volume (the interior) of the rigid body and $\partial V$ its boundary (its surface). Denoting by $\mu(\bx)$ the classical mass density of the rigid body, we have that
\begin{multline}
\dot{E}^{\rm CoM}_t\vert_{V}
=\\=
\frac{\hbar^2}{2 M} \sum_{A=C,G} \int_{V} \dd[3]{\bx}\dd[3]{\by} \gamma_A (\bx-\by) \prtq{\nabla \mu (\bx)}\cdot \prtq{\nabla \mu (\by)},
\end{multline}
which is independent of the form of the smearing distribution. If the body has constant density, this term gives a vanishing contribution. Denoting by $\mu_S$ the mass density near the surface of the rigid body and assuming it to be constant, we have that
\begin{multline}
\dot{E}^{\rm CoM}_t\vert_{\partial V}
=\\=
\frac{\hbar^2 (2\pi)^{9/2} \mu_S^2 }{2 M} \sum_{A=C,G} 
\int \dd[3]{\bk} \bk^2 \tl{\gamma}_A (\bk) \abs{\tl{\chi}_V(\bk) \tl{g}_{r_A}(\bk)}^2,
\end{multline}
where $\chi_V (\bx)$ is the indicator function associated to the shape of the rigid body and we recall that $\tl{f}$ denotes the Fourier transform of $f$. It does not seem possible to reduce the above quantity to an expression proportional to $I_{\gamma_C}[g_{r_C}] + I_{\gamma_G}[g_{r_G}]$, as we did for the radiation. Thus, we conclude that optimizing over the diffusion of the center of mass of a large rigid body probably leads to a different smearing profile than the one we found for minimizing the \emph{total} energy increase.

As shown in Table~\ref{tab:my-table}, replacing the optimal profile with a Gaussian increases the predicted diffusion (at fixed $r_C$) by $47\%$ for CSL and $22\%$ for DP (and TD). Thus, while the impact of using the optimal smearing is by no means negligible, our results also show that, until experiments reach much higher precision~\cite{Carlesso2022Present,Janse2024BoundsSpacetimeDiffusion,Piccione2025ExploringMassDependence}, Gaussian distributions are practically interchangeable with the optimal distributions. As experimental sensitivity improves, however, this may not hold. In particular for the TD model, for which both upper and lower bounds are available from experiments~\cite{Tilloy2016CSLGravity}, the Gaussian-based version may be ruled out while the version based on the optimal distribution may not be. The converse seems less probable.
This is especially true when considering that the strongest bounds on the DP and TD models come from the prediction of spontaneous emission of radiation~\cite{XENONCollaboration2026CollapseConstraints}, which we have shown to be deeply connected with spontaneous diffusion.

\section{Discussion\label{Sec:Discussion}}

Our proposal (fix $r_C$, and possibly $r_G$, and pick the smearing(s) that minimize the heating rate) allows us to remove an otherwise ad-hoc modeling choice of collapse and hybrid models while keeping the large-scale phenomenology intact. In this sense, for each $r_C$ (and possibly $r_G$), it singles out the most conservative variant whose experimental refutation would also disfavor any sub-optimal profile. What it selects, however, depends on the constraints: here we required non-negativity, normalization, and fixed variance of the smearing distribution(s); choices we view as the most natural ones. Different constraints would, in most cases, change the optimizer(s).

We now comment on the choice of identifying the smearing length $r_C$ through the smearing distribution's variance [see Eq.~\eqref{eq:SmearingConstraints}].
The variance provides a natural criterion because, once normalization is imposed (zeroth moment constraint) and the distribution is centered (vanishing first moment constraint), the second moment (i.e., the variance) is the lowest-order nontrivial moment that quantifies its spatial spread. In other words, fixing the variance amounts to imposing the minimal moment constraint needed to compare different smearings at the same length scale, while still leaving their shape otherwise unconstrained.
Moreover, the use of the variance also admits a simple operational interpretation. By Markov's inequality~\cite[page 13]{Book_Bhattacharya2016BasicCourseProbabilityTheory} one has
\begin{equation}
\int_{\frac{\abs{\bx}}{r_C}\geq \kappa}
\!\!\!\!\!\!\!\!
\dd[3]{\bx} g_{r_C}(\bx) \leq \frac{3 r_C^2}{\kappa^2 r_C^2} = \frac{3}{\kappa^2},
\quad
\kappa>0.
\end{equation}
Thus, fixing the variance bounds how much of the smearing distribution can lie arbitrarily far from the origin, showing that $r_C$ really characterizes the spatial spread of the smearing.

The claim that the large-scale phenomenology of collapse and hybrid models is independent of the smearing distribution choice seems in contrast with the long-range (superposition distances much higher than $r_C$) decoherence (or collapse) rate of a single particle for models like CSL, DP, and TD. For example, with a Gaussian smearing one has that the long-range collapse rate of CSL is given by $\Gamma \simeq \gamma_{\rm CSL} (4 \pi r_C^2)^{3/2}$~\cite{Review_Bassi2013Models} (see also Appendix~\ref{APPSec:CSL_Appendix}). However, if one considers the collapse rate of a macroscopic rigid body, one gets a rate independent of both $r_C$ and the smearing distribution. We show that this is the case for all such models in Appendix~\ref{APPSec:WeaklyContinuousMonitoring}.
In passing, we notice that this same calculation with hybrid models (see Appendix~\ref{APPSubsec:WeaklyContinuousMonitoring_DecoherenceRate}) implies that applying the PLD of Ref.~\cite{Tilloy2017LeastDecoherence} to large macroscopic bodies yields the DP correlator without assuming that the measurement and feedback smearings have to be equal.

It may come as a surprise that GRW and CSL yield different optimal smearings, as their single-particle master equations have the same operator form and the heating rate is a sum of single particles terms. However, the difference arises because the functional to minimize is different, i.e., $I[\sqrt{g_{r_C}}]$ in GRW vs. $I[g_{r_C}]$ in CSL (see Table~\ref{tab:my-table}).
It is also worth noting that choosing $g_{r_C}$ to be Gaussian in both models makes an exact mapping possible for the single particle dynamics, by virtue of the fact that the square root of a Gaussian is another (non-normalized) Gaussian. This is not so for a generic $g_{r_C}$ (see Appendix~\ref{APPSec:CSL_Appendix} for an example).

To finish our discussion, let us notice that CSL and DP can be equivalently considered as part of a broader translation-covariant framework in which the mass density is not smeared: one may instead shape the spatial correlator $\gamma_C (\bx,\by)$ so that the heating remains finite~\cite{Review_Bassi2013Models}. Both CSL and DP can become of this type by simply transferring the smearing from the mass density operator on to the correlator, i.e., $\gamma_C (\bx,\by) \rightarrow (g_{r_C}*\gamma_C*g_{r_C}) (\bx,\by)$ [cf. Eqs.~\eqref{eq:HybridModelsEquation} and~\eqref{eq:HybridModelsEquation_Components}]. However, the contrary is not true. For example, one could allow for a $\gamma_C (\bx,\by)$ that also takes negative values. This may have physical relevance because one could argue, for example, that the CSL dynamics may come from the interaction of quantum matter with a stochastic classical field being delta-correlated in time and characterized by Gaussian correlations in space~\cite{Review_Bassi2013Models}. However, exploring this larger model space is beyond the scope of this paper but may be interesting for future work.

\section{Conclusions\label{Sec:Conclusions}}

We proposed a simple procedure for choosing the spatial smearing used in collapse and hybrid classical–quantum gravitational models: given the distribution variance $r_C^2$, one should search for the distribution that minimizes the heating rate. This criterion keeps other large-scale predictions intact while removing an otherwise arbitrary modeling choice. We argued that this identifies, for a given model and smearing length $r_C$, the most conservative variant with respect to spontaneous diffusion and, therefore, the minimal deviation from standard quantum mechanics.

We applied our procedure to the most investigated collapse models (GRW, CSL, DP), singling out a unique profile for the smearing distribution in each case. Then, we applied it to the TD hybrid gravitational model with matched smearings (measurement and feedback). This leads to a hybrid model of semi-classical gravity that is entirely characterized by a single parameter $r_C$, which can be experimentally bounded from above and below~\cite{Tilloy2016CSLGravity,Tilloy2017LeastDecoherence}. When the two smearings differ, we showed how one gets two separate state-independent optimization problems.
We remark, however, that our procedure can be applied to any collapse or hybrid models which can be seen as based on continuous weak measurements of the (smeared) mass density operator.

The most important phenomenological deviation from standard quantum mechanics due to collapse and hybrid models is the breakdown of the superposition principle. As we argued, however, its manifestation is independent of the chosen smearing functions at macroscopic scales.
Energy non-conservation constitutes another prominent, directly testable phenomenological deviation from standard quantum mechanics, which we propose to minimize while keeping $r_C$ fixed. Thus, our proposal reduces the functional freedom of these models to a small set of parameters. Experimentally ruling out all possible values of them, especially for the hybrid gravitational models, would arguably rule out the entire class of models, independently of the smearing.

\section*{Acknowledgements}

N.P. thanks A. Bassi for useful discussions about the optimization results on the CSL model. 
N.P. is thankful to people attending the conference ``A look at the interface between gravity and quantum theory – 2025 edition'', in particular A. Kent, L. Diosi, and A. Di Biagio; discussions at the workshop have inspired this work.
N.P. thanks A. Di Biagio and A. Bassi for numerous and useful comments about the draft of this work.

N. P. acknowledges support from INFN and the University of Trieste.
N. P. acknowledges support also from the European Union Horizon’s 2023 research and innovation programme [HORIZON-MSCA-2023-PF-01] under the Marie Sk\l{}odowska Curie Grant Agreement No. 101150889 (CPQM).

%\bibliography{Bibliography}


\clearpage
\onecolumngrid
\appendix

\section{Minimization of the GRW heating rate\label{APPSec:GRW_Appendix}}

In the GRW model, the master equation for a generic system of $N$ particles takes the form~\cite{Review_Bassi2003Dynamical}
\begin{equation}\label{APPeq:GRW_MasterEquation}
\dv{t}\rho_t = -\frac{i}{\hbar}\comm{\hH}{\rho_t} -\sum_j \lambda_j\prtg{\rho_t - \int \dd[3]{\bx} \sqrt{g_{r_C} (\bx-\hbq_j)}\rho_t \sqrt{g_{r_C} (\bx-\hbq_j)}},
\end{equation}
and the energy-increase rate is given by
\begin{equation}
\dot{E}_t = \hbar^2 \prtq{\sum_j \frac{\lambda_j}{m_j}} I[\sqrt{g_{r_C}}],
\end{equation}
where $I[f]$ is the so-called Dirichlet energy of the function $f$~\cite{Book_Evans2010PartialDifferentialEquations}, i.e.:
\begin{equation}\label{APPeq:FunctionalDefinition}
I[f]
= \frac{1}{2}\int \dd[3]{\bx} \abs{\nabla f(\bx)}^2
= \frac{1}{2}\int \dd[3]{\bk} \bk^2 \abs{\tl{f}(\bk)}^2,
\end{equation}
with $\tl{f} (\bk) = (2\pi)^{-3/2} \int \dd[3]{\bx} f(\bx) e^{-i \bk \cdot \bx}$. 

To minimize the energy increase it is sufficient to minimize $I[\sqrt{g_{r_C}}]$ under the constraints of Eq.~\eqref{eq:SmearingConstraints}. This is an already solved problem as it reduces to the problem of saturating the three-dimensional uncertainty principle for the ``wavefunction'' $\psi (\bx) = \sqrt{g_{r_C} (\bx)}$~\cite{Folland1997SurveyUncertaintyPrinciple,McCurdy2021StabilityHPWInequality,Fathi2021ShortProofHPWInequality,Review_Dodonov2025UncertaintyRelations}. The solution is given by Gaussian wavefunctions~\cite{McCurdy2021StabilityHPWInequality,Fathi2021ShortProofHPWInequality}.

We can compare the value of $I[f]$ using the optimal distributions derived for the CSL (Appendix~\ref{APPSec:CSL_Appendix}) and DP (Appendix~\ref{APPSec:DP_Appendix}) models:
\begin{equations}
	&\textrm{GRW:}\ &
	&f_{r_C}(\bx) = \sqrt{\frac{\exp{-\bx^2/(2 r_C^2)}}{(2 \pi r_C^2)^{3/2}}}&
	&\implies&
	&I[f_{r_C}] = \frac{3}{8 r_C^2}
	= 0.375 \times r_C^{-2}.&
	\\
	&\textrm{CSL:}\ &
	&f_{r_C}(\bx) = \sqrt{\frac{105}{32 \pi R^7}\prt{R^2-\bx^2}^2 \Theta(R - \abs{\bx})}\vert_{R=3 r_C}&
	&\implies&
	&I[f_{r_C}] = \frac{7}{12 r_C^2}
	\simeq 0.583 \times r_C^{-2}.&
	\\
	&\textrm{DP:}\ &
	&f_{r_C} (\bx) = \sqrt{\frac{15}{8 \pi R^5}\prt{R^2 - \bx^2}\Theta(R - \abs{\bx})}\vert_{R=\sqrt{7}r_C}&
	&\implies&
	&I[f_{r_C}] = +\infty.&
\end{equations}
Indeed, the Gaussian distribution gives the lowest value.

\clearpage
\section{Calculations for the CSL model\label{APPSec:CSL_Appendix}}

\subsection{Minimization of the CSL heating rate\label{APPSubsec:CSL_MinimizationProblem}}

To simplify the notation, here we set $r_C=1$ and we write $g$ in place of $g_{r_C}$. 

We want to minimize the quantity
\begin{equation}
I[g]
:= \frac{1}{2}\int \dd[3]{\bx} [\nabla g (\bx)]\cdot[\nabla g (\bx)]
= \frac{1}{2}\int \dd[3]{\bk} \bk^2 \abs{\tl{g} (\bk)}^2,
\qquad
\tl{g} (\bk) = (2\pi)^{-3/2} \int \dd[3]{\bx} g(\bx) e^{-i \bk \cdot \bx},
\end{equation}
under the constraints
\begin{equation}\label{APPeq:SmearingConstraints}
\int \dd[3]{\bx} g (\bx)  = 1,
\qquad
\int \dd[3]{\bx} \bx^2 g (\bx)  = 3,
\qquad
g(\bx) \geq 0.
\end{equation}
First, we notice that the functional $I$ is convex ($\lambda \in [0,1]$, $f$ and $h$ are generic functions):
\begin{equation}
I[\lambda f + (1-\lambda)h]
=
\frac{1}{2}\int \dd[3]{\bk} \bk^2 \abs{\lambda \tl{f} + (1-\lambda)\tl{h} }^2
\leq
\frac{1}{2}\int \dd[3]{\bk} \bk^2 \prtq{\lambda \abs{\tl{f}}^2 + (1-\lambda)\abs{\tl{h}}^2}
=
\lambda I[f] + (1-\lambda)I[h].
\end{equation}
This means that any local minimum in the space of allowed functions is also a global minimum~\cite[see Theorem 1.2.2]{Book_BorweinVanderwerff2010ConvexFunctions}. Indeed, the space of allowed functions is itself convex\footnote{In particular and more precisely, we assume that $f \in H^{1}$, where $H^1$ is the space of functions in $L^2(\mathbb{R})$ such that also $f'\in L^2(\mathbb{R})$. So, $H^1$ is a Banach space and the restriction to such a set given by the constraints gives a convex subset of $H^1$. Since we find the minimum element in this set, it is the global minimum also in $L^2(\mathbb{R})$ (also a Banach space).}:
\begin{equation}
\int \dd[3]{\bx} \prt{\lambda f (\bx) + (1-\lambda)h (\bx)}  = 1,
\qquad
\int \dd[3]{\bx} \bx^2 \prt{\lambda f (\bx) + (1-\lambda)h (\bx)}  = 3,
\qquad
\lambda f (\bx) + (1-\lambda)h (\bx) \geq 0.
\end{equation}

The Pólya–Szegő inequality for symmetric decreasing rearrangements~(Theorem 2.3.1 in Ref.~\cite{Book_Kesavan2006Symmetrization}) assures us that $g(\bx)$ has to be radial and decreasing. In fact, let us consider a generic $g(\bx)$. Then, we would have\footnote{The general Pólya–Szegő inequality gives $\norm{\nabla g^*}_{p}\leq \norm{\nabla g}_{p}$.}
\begin{equation}
\norm{g^*}_{1} = \norm{g}_{1},
\quad
\norm{\bx^2 g^*}_1 \leq \norm{\bx^2 g}_1,
\quad 
\norm{\nabla g^*}_{2}\leq \norm{\nabla g}_{2},
\qquad
\norm{g}_{p} := \prt{\int \dd[3]{\bx} \abs{g(\bx)}^p}^{1/p},
\end{equation}
where $g^*(\bx)$ denotes the symmetric decreasing rearrangement, that is, 
denoting by $\Omega$ any measurable set, and by $\Omega^*$ the ball centered at the origin with the same volume, one defines $g^* (\bx) = \int_0^{\infty} \chi^*_{\prtg{\abs{g}>t}} (\bx) \dd{t}$, where $\chi^*_{\Omega} = \chi_{\Omega^*}$ and $\chi_{\Omega}$ is the indicator function associated to the set $\Omega$ (see Chapter 3 of Ref.~\cite{Book_Lieb2001Analysis} or Chapter 1 of Ref.~\cite{Book_Kesavan2006Symmetrization}). Now, suppose that the optimal $g$ is not radial and decreasing. Since $I[g]=(1/2)\norm{\nabla g}_{2}$, it follows that $I[g^*]\leq I[g]$, although $g^* (\bx)$ does not satisfy the variance constraint, i.e., $\int \dd[3]{\bx} \bx^2 g (\bx)  = 3 \alpha^2$, with $0 < \alpha < 1$. Then, we may define the new function $h^* (\bx) = \alpha^3 g^*(\alpha \bx)$, which is still normalized, has the correct variance, and gives $I[h^*]=\alpha^5 I[g^*] < I[g]$. This contradiction proves that the optimal distribution $g (\bx)$ has to be decreasing.

Since we have shown that $g(\bx)$ is a radial distribution, writing $r=\abs{x}$, we can now consider the one-dimensional version of the problem:
\begin{equation}
	I[g]= 2\pi \int_0^\infty \dd{r} r^2 [g'(r)]^2,
	\qquad
	4\pi \int_0^\infty \dd{r} r^2 g(r) =1,
	\quad
	4\pi \int_0^\infty \dd{r} r^4 g(r) =3,
	\quad
	g(r) \geq 0.
\end{equation}
Moreover, since $g(r)$ is decreasing and non-negative, its support is $[0,R)$, with $R>0$ or $R=+\infty$. 
%Moreover, since the quantity to minimize involves the first derivative of $g(r)$, the minimizer function must be derivable everywhere~[Refs].
We assume that the minimizer function is derivable everywhere, which restricts the set of allowed function to a smaller but still convex subset. It follows
%So, we assume the first derivative of $g(r)$ to exist everywhere so 
that $g'(R)=0$, because for $r>R \implies g(r)=0$ (for the derivative to exist, left and right derivatives must coincide). Finally, since $g(r)$ is radial, its derivative in $r=0$ must be zero.

The optimization problem can be recast with Lagrangian multipliers:
\begin{equation}
\mcL[g] = 4\pi \int_0^{R} \prt{\frac{1}{2}r^2 [g'(r)]^2 + \lambda r^2 g(r) + \mu r^4 g(r)}\dd{r} - \lambda - 3\mu.
\end{equation}
Assuming that $g$ is the minimizing function, we demand that $\dv{\varepsilon}\mcL[g+\varepsilon \phi]\vert_{\varepsilon=0}=0$. We get
\begin{equation}
\dv{\varepsilon}\mcL[g+\varepsilon \phi]\vert_{\varepsilon=0} = \int_0^R \dd{r}\prtg{r^2 g' \phi' + \lambda r^2 \phi + \mu r^4 \phi}=0.
\end{equation}
Integrating by parts and using the fact that $g'(0)=g'(R)=0$, the above formula becomes
\begin{equation}
\int_0^R \dd{r}\prtg{- 2 r g' - r^2 g'' + \lambda r^2 + \mu r^4}\phi=0,
\implies
2 \frac{g'(r)}{r} + g''(r) = \lambda + \mu r^2.
\end{equation}
With the condition $g'(0)=0$, the above differential equation has solution
\begin{equation}
g(r) = c + \frac{\lambda}{6}r^2 + \frac{\mu}{20}r^4,
\quad
g'(r) = \frac{\lambda}{3}r + \frac{\mu}{5}r^3.
\end{equation}
Imposing the conditions $g(R)=g'(R)=0$, one gets
\begin{equation}
\lambda = -\frac{3}{5}\mu R^2,
\quad
\mu = 20 c R^{-4}
\implies
g(r) = \frac{c}{R^4}\prt{R^2-r^2}^2.
\end{equation}
To fix $c$ and $R$, we use the normalization and variance constraints:
\begin{equation}
4 \pi \int_0^R r^2 g(r) = c \frac{32 \pi}{105}R^3 = 1,
\quad
4 \pi \int_0^R r^4 g(r) = c \frac{32 \pi}{315}R^5 = 3,
\implies
R=3,\quad c= \frac{105}{32 \pi R^3}.
\end{equation}
Restoring the physical dimensions of $r_C$ one has that $R=3 r_C$ and the equation reported in Table~\ref{tab:my-table} follows.

The optimal distribution for the CSL model, as a single-variable function of $r=\abs{\bx}$, does not have a second derivative in $r=3 r_C$, and is a compact-support function. However, both these potential sources of problems can be cured by convoluting it with a Gaussian of arbitrarily small radius $\varepsilon$. Thus, if one searches for an optimal distribution which is smooth and without compact support, there would be no solution\footnote{The value $I[g_{r_C}]$ computed for the optimal distribution for the CSL model constitutes the infimum of the image of the functional $I[f]$ with domain given by the set of smooth and positive functions satisfying the constraints.}.

To give some numerical examples, we can compare the value of $I[g]$ using the optimal distributions derived for the GRW (Appendix~\ref{APPSec:GRW_Appendix}) and DP (Appendix~\ref{APPSec:DP_Appendix}) models:
\begin{equations}
&\textrm{GRW:}\ &
&g_{r_C}(\bx) = \frac{\exp{-\bx^2/(2 r_C^2)}}{(2 \pi r_C^2)^{3/2}}&
&\implies&
&I[g_{r_C}] = \frac{3}{32 \pi^{3/2} r_C^5}
\simeq 0.0168 \times r_C^{-5}.&
\\
&\textrm{CSL:}\ &
&g_{r_C}(\bx) = \frac{105}{32 \pi R^7}\prt{R^2-\bx^2}^2 \Theta(R - \abs{\bx})\vert_{R=3 r_C}&
&\implies&
&I[g_{r_C}] = \frac{35}{972 \pi r_C^5}
\simeq 0.0115 \times r_C^{-5}.&
\\
&\textrm{DP:}\ &
&g_{r_C}(\bx) = \frac{15}{8 \pi R^5}\prt{R^2-\bx^2} \Theta(R - \abs{\bx})\vert_{R=\sqrt{7}r_C}&
&\implies&
&I[g_{r_C}] = \frac{45}{392 \sqrt{7} \pi r_C^5}
\simeq 0.0138 \times r_C^{-5}.&
\end{equations}
Using the Gaussian in place of the optimal distribution increases the Dirichlet energy by roughly $47\%$.

\subsection{Long-range collapse rate of the CSL model\label{APPSubsec:CSL_LongRangeDecoherence}}

Let us first consider a single particle of mass $m$ in the CSL master equation. We show that by opportunely choosing $\gamma_{\rm CSL}$ one can always get the desired long-range collapse rate. Since we are interested in the collapse rate, in this case we can ignore the standard Hamiltonian of the system
\begin{equation}
\dot{\rho}_t 
=
-\frac{\gamma_{\rm CSL}}{2} \prt{\frac{m}{m_0}}^2 \int \dd[3]{\bx} \comm{g_{r_C} (\bx-\hbq)}{\comm{g_{r_C} (\bx-\hbq)}{\rho_t}}
=
-\gamma_{\rm CSL} \prt{\frac{m}{m_0}}^2 \prtq{K \rho_t - \int \dd[3]{\bx} g_{r_C} (\bx-\hbq) \rho_t g_{r_C} (\bx-\hbq)},
\end{equation}
where $K:= \int \dd[3]{\bx} g^2_{r_C} (\bx)$. 
In position representation, one gets that
\begin{equation}
\dot{\rho}_t (\bx,\by) =
-\Gamma(\bx-\by)\rho_t (\bx,\by),
\qq{where}
\Gamma (\bd) =
\gamma_{\rm CSL}\prt{\frac{m}{m_0}}^2 \prtq{K - \int \dd[3]{\bx} g_{r_C} (\bx)g_{r_C} (\bx+\bd)}.
\end{equation}
Now, considering a superposition at distances much larger than $r_C$, the last term in $\Gamma (\bd)$ goes to zero and one gets
\begin{equation}
\Gamma (\bd) =
\gamma_{\rm CSL}\prt{\frac{m}{m_0}}^2 \prtq{K - \int \dd[3]{\bx} g_{r_C} (\bx)g_{r_C} (\bx+\bd)}
\rightarrow
\gamma_{\rm CSL} K \prt{\frac{m}{m_0}}^2
=
\lambda_{\rm CSL} \prt{\frac{m}{m_0}}^2.
\end{equation}
So, whatever the value of $K>0$ may be, one can always choose $\gamma_{\rm CSL}$ in such a way to obtain the collapse rate $\lambda_{\rm CSL}$. When $g_{r_C} (\bx)$ is a Gaussian distribution, this gives the usual factor
\begin{equation}
g_{r_C} (\bx) = \frac{e^{-\bx^2/2 r_C^2}}{(2\pi r_C^2)^{3/2}}
\implies
K = \int \dd[3]{\bx} g^2_{r_C} (\bx) = (2 \sqrt{\pi} r_C)^{-3}
\implies
\lambda_{\rm CSL} = \frac{\gamma_{\rm CSL}}{(4 \pi r_C^2)^{3/2}}.
\end{equation}
On the other hand, using the optimal distribution for the CSL model one gets
\begin{equation}
g_{r_C} (\bx) = \frac{105}{32 \pi (3 r_C)^7}\prtq{9 r_C^2-\bx^2}_+^2
\implies
K = \int \dd[3]{\bx} g^2_{r_C} (\bx) = \frac{35}{594 \pi r_C^3}
\implies
\lambda_{\rm CSL} = \gamma_{\rm CSL} \frac{35}{594 \pi r_C^3}.
\end{equation}

When considering many particles, the situation becomes more complicated. However, if we focus on a rigid body whose density varies on scales much larger than the collapse radius $r_C$, the collapse rate of the body's CoM can be approximated as follows~\cite[see Eq.~(25)]{Piccione2025ExploringMassDependence}:
\begin{equation}
\Gamma(\bd) \simeq \frac{\gamma_{\rm CSL}}{m_0^2}\int \dd[3]{\bx} 
\prtg{\mu^2 (\bx) - \mu(\bx)\mu(\bx+\bd)},
\qquad
\Gamma(\infty) \simeq \frac{\gamma_{\rm CSL}}{m_0^2}\int \dd[3]{\bx} 
\mu^2 (\bx),
\end{equation}
where $\mu (\bx)$ is the body's mass density. Indeed, the result is independent of the smearing distribution and, this time, the collapse rate is not given by $\lambda_{\rm CSL}$ but it is directly proportional to $\gamma_{\rm CSL}$.

\subsection{Comparison between GRW and CSL for a single particle\label{APPSubsec:CSL_ComparisonGRWCSL}}

When dealing with a single particle, the CSL and GRW models are functionally equivalent. The collapse rate of GRW and CSL read
\begin{equations}
\Gamma_{\rm GRW} (d) 
&= 
\lambda_{\rm GRW}\prt{\frac{m}{m_0}}\prtq{1-\int \dd[3]{\bx} \sqrt{g_{r_C} (\bx)} \sqrt{g_{r_C} (\bx+\bd)}},
\\
\Gamma_{\rm CSL} (d) 
&= 
\lambda_{\rm CSL}\prt{\frac{m}{m_0}}^2\prtq{1-\frac{1}{K}\int \dd[3]{\bx} g_{r_C} (\bx) g_{r_C} (\bx+\bd)}.
\end{equations}
When using a Gaussian smearing the two can be made exactly equivalent because~\cite{Piccione2025ExploringMassDependence} (we now take $m=m_0$ to simplify the formulas)
\begin{equation}
g_{r_C} (\bx) = \frac{e^{-\bx^2/2 r_C^2}}{(2\pi r_C^2)^{3/2}}
\implies
\Gamma_{\rm GRW} (d) = \lambda_{\rm GRW}\prtq{1-e^{-d^2/8 r_C^2}},
\qquad
\Gamma_{\rm CSL} (d) = \lambda_{\rm CSL}\prtq{1-e^{-d^2/4 r_C^2}}.
\end{equation}
Indeed, defining $r_C' = r_C/\sqrt{2}$ and $\lambda_{\rm GRW}=\lambda_{\rm CSL}$ in the CSL case, one gets the exact same thing.
If, instead, we use the optimal distribution that we found for the CSL model, we get (with $R=3 r_C$)
\begin{equation}
g_{r_C} (\bx) = \frac{105}{32 \pi R^7}\prt{R^2-\bx^2}^2 \Theta(R - \abs{\bx})
\implies
\Gamma_{\rm GRW} (d) = \lambda_{\rm GRW}\prtq{1-F_{\rm GRW}(s)},
\qquad
\Gamma_{\rm CSL} (d) = \lambda_{\rm CSL}\prtq{1-F_{\rm CSL}(s)}
\end{equation}
where $s:= d/R$ and
\begin{equations}
F_{\rm GRW}(s) &:= \prt{1-\frac{7}{4}s^2 + \frac{35}{32}s^3 - \frac{7}{64}s^5 + \frac{3}{512}s^7}\Theta (2-s),
\\
F_{\rm CSL}(s) &:= \prt{1-\frac{11}{6}s^2+\frac{33}{16}s^4-\frac{77}{64}s^5+\frac{33}{256}s^7-\frac{11}{1024}s^9+\frac{5}{12288}s^{11}}\Theta (2-s).
\end{equations}
This result shows that the equivalence between CSL and GRW for a single particle is a byproduct of choosing the Gaussian smearing. Indeed, the square root of a normalized Gaussian is another (non-normalized) Gaussian, but this is not so for other distributions.

\clearpage
\section{Minimization of the DP heating rate\label{APPSec:DP_Appendix}}

For the minimization of the DP heating rate, we follow the same strategy adopted in the CSL case. In this appendix, we set $r_C=1$ to lighten the notation.

We want to minimize the quantity
\begin{equation}\label{APPeq:DP_FunctionalToMinimize}
I_{\rm DP}[g]
:= \frac{1}{4}\int \dd[3]{\bx} \dd[3]{\by} \frac{1}{\abs{\bx-\by}}[\nabla g (\bx)]\cdot[\nabla g (\by)]
= \pi \int \dd[3]{\bk} \abs{\tl{g} (\bk)}^2
= \pi \int \dd[3]{\bx} \abs{g (\bx)}^2,
\end{equation}
under the constraints
\begin{equation}
\int \dd[3]{\bx} g (\bx)  = 1,
\qquad
\int \dd[3]{\bx} \bx^2 g (\bx)  = 3,
\qquad
g(\bx) \geq 0.
\end{equation}

Similar arguments to those used for CSL show that $g(r)$ has to be radial and decreasing\footnote{For any function $g$, its decreasing rearrangement $g^*$ satisfies $\norm{g^*}_p=\norm{g}_p$, where $p=1$ gives the normalization condition and $I_{\rm DP}[g]\propto \norm{g}_2$. The variance of $g^*$ may be lower, i.e., $\norm{\bx^2 g^*}_1=3 \alpha^2$ with $0<\alpha\leq 1$. Then, using $h(\bx)=\alpha^3 g^*(\alpha \bx)$ leads to $\norm{h}_1=\norm{g}_1$ and $\norm{\bx^2 h}_1 = 3$, but also $\norm{h}_2 = \alpha^{3/2} \norm{g^*}_2 \leq \norm{g}_2$. Thus, a not radial and decreasing $g(\bx)$ cannot be the minimizer.}. So, writing $r=\abs{x}$, we can now consider the one-dimensional version of the problem:
\begin{equation}
	I_{\rm DP}[g]= 4\pi^2 \int_0^\infty \dd{r} r^2 [g(r)]^2,
	\qquad
	4\pi \int_0^\infty \dd{r} r^2 g(r) =1,
	\quad
	4\pi \int_0^\infty \dd{r} r^4 g(r) =3,
	\quad
	g(r) \geq 0.
\end{equation}
The support of $g(r)$ is $[0,R)$, with $R>0$ or $R=\infty$, with $g(R)=0$. The quantity to minimize (once rewritten as in the rightmost equality of Eq.~\eqref{APPeq:DP_FunctionalToMinimize}) does not involve derivatives now, so we do not assume that $g(r)$ is derivable everywhere. In fact, we will find a minimizer that is not derivable in $r=R$.

The optimization problem can be recast with Lagrangian multipliers:
\begin{equation}
\mcL[g] = 4\pi \int_0^{R} \prt{\pi r^2 [g(r)]^2 + \lambda r^2 g(r) + \mu r^4 g(r)}\dd{r} - \lambda - 3\mu.
\end{equation}
Assuming that $g$ is the minimizing function, we demand that $\dv{\varepsilon}\mcL[g+\varepsilon \phi]\vert_{\varepsilon=0}=0$. We get
\begin{equation}
\dv{\varepsilon}\mcL[g+\varepsilon \phi]\vert_{\varepsilon=0} = \int_0^R \dd{r}\prtg{2 \pi r^2 g + \lambda r^2 + \mu r^4 }\phi=0,
\implies
2\pi g + \lambda + r^2 \mu = 0.
\end{equation}
From the above, we get that
\begin{equation}
g(r) = - \frac{\lambda + \mu r^2}{2 \pi},
\qquad
g(R)=0 
\implies \lambda = -\mu R^2
\implies g(r) = \frac{\mu}{2\pi}\prt{R^2 -r^2}.
\end{equation}
Finally, using the constraints we get
\begin{equation}
1
=4\pi \int_0^{R} \dd{r} r^2 g(r)
= \frac{\mu R^5}{15}
\implies
\mu = 15 R^{-5},
\qquad
3 
= 4\pi \int_0^{R} \dd{r} r^4 g(r)
= \frac{3}{7}R^2,
\implies
R = \sqrt{7}.
\end{equation}
Restoring the physical units, one gets the result reported in Table~\ref{tab:my-table}.

A numerical comparison between using the optimal distribution and the Gaussian one gives
\begin{equations}
&\textrm{GRW:}\ &
&g_{r_C}(\bx) = \frac{\exp{-\bx^2/(2 r_C^2)}}{(2 \pi r_C^2)^{3/2}}&
&\implies&
&I_{\rm DP}[g_{r_C}] = \frac{1}{8 \pi^{1/2} r_C^3}
\simeq 0.0705 \times r_C^{-3}.&
\\
&\textrm{CSL:}\ &
&g_{r_C}(\bx) = \frac{105}{32 \pi R^7}\prt{R^2-\bx^2}^2 \Theta(R - \abs{\bx})\vert_{R=3 r_C}&
&\implies&
&I_{\rm DP}[g_{r_C}] = \frac{35}{22 (3 r_C)^3}
\simeq 0.0589 \times r_C^{-3}.&
\\
&\textrm{DP:}\ &
&g_{r_C} (\bx) = \frac{15}{8 \pi R^5}\prt{R^2 - \bx^2}\Theta(R - \abs{\bx})\vert_{R=\sqrt{7}r_C}&
&\implies&
&I_{\rm DP}[g_{r_C}] = \frac{15}{14  (\sqrt{7} r_C)^3}
\simeq 0.0579 \times r_C^{-3}.&
\end{equations}
Using the Gaussian in place of the optimal distribution increases the value of the functional $I_{\rm DP} [g_{r_C}]$ by roughly $22\%$.

\clearpage
\section{Models based on continuous weak measurements of the mass density
\label{APPSec:WeaklyContinuousMonitoring}}

In this appendix, we will review the approach to continuous weak monitoring models following Refs.~\cite{Tilloy2016CSLGravity,Tilloy2017LeastDecoherence,GaonaReyes2021GravitationalFeedback}\footnote{The first two subsections of this Appendix are very similar to those of Appendix A in Ref.~\cite{Piccione2025NewtonianPSL}, the main difference being that here we do not assume that the smearing procedure for the measurement and feedback parts of the model are the same.}. Weak monitoring models of hybrid Newtonian gravity are constructed by considering naturally occurring continuous weak measurements of the smeared mass density\footnote{We recall that the smeared mass density operator is defined by $\hmu_{r_C} (\bx) = (g_{r_C} * \hmu) (\bx)$ where $g_{r_C} (\bx)$ is a smearing function usually characterized by a radius $r_C$ and the $*$ operator denotes convolution. Usually, one takes the Gaussian smearing $g_{r_C}(\bx) = (2 \pi r_C^2)^{-3/2}\exp{-\bx^2/(2 r_C^2)}$. This smearing is necessary to avoid divergences and $r_C$ usually constitutes a free parameter of spontaneous collapse models.} $\hmu_{r_C} (\bx)$ at all points in space, and implementing Newtonian gravity as a feedback mechanism based on the measurements' results. For future convenience, we also introduce the notation $\tmuR (\bx) = \hmu_{r_C} (\bx) - \ev{\hmu_{r_C} (\bx)}$. 

In the first subsection, we will only deal with the measurement part, thus obtaining a general collapse model that includes CSL and DP. 
In the second subsection, we will add the gravitational feedback so that we get a general class of hybrid gravitational models, i.e., those considered in Ref.~\cite{Tilloy2016CSLGravity}.
In the third subsection, we compute the spatial decoherence rate for a single particle and the Center of Mass (CoM) of a rigid body. We also show that the result of Ref.~\cite{Tilloy2017LeastDecoherence} does not generally hold without the assumption of equal smearing in measurement and feedback parts.
Finally, in the fourth subsection, we compute the approximated diffusion of the CoM of a rigid body.

\subsection{The Measurement Part\label{APPSubsec:WeaklyContinuousMonitoring_MeasurementPart}}

We consider that, at each spatial point, a continuous weak measurement of the smeared mass density operator takes place, giving the measurement record~\cite{GaonaReyes2021GravitationalFeedback}:
\begin{equation}
	\mu_t (\bx) = \ev{\hmu_{r_C} (\bx)} + \delta \mu_t (\bx),
	\qquad
	\delta \mu_t (\bx) = \frac{1}{2}\int \dd[3]{\by} \gamma_C^{-1} (\bx-\by)\frac{\dd{W_t} (\by)}{\dd{t}},
\end{equation}
where the Wiener increment $\dd{W_t} (\bx)$ is such that\footnote{This kind of generalized Wiener increment can be obtained by convoluting the usual Wiener increment. See page 22 (492) of Ref.~\cite{Review_Bassi2013Models}.} $\dd{W_t} (\bx) \dd{W_t} (\by) = \gamma_C (\bx,\by) \dd{t} = \gamma_C \prt{\abs{\bx-\by}} \dd{t}$. 
Moreover, one has $\dd{W_t} (\bx) \dd{t} =0$ and $\mbE [\dd{W_t} (\bx)] = 0$.
The inverse of $\gamma_C (\bx,\by)$ is defined by the relation 
\begin{equation}
	(\gamma_C \circ \gamma_C^{-1} )(\bx,\by) \equiv 
	\int \dd[3]{\bz} \gamma_C (\bx,\bz)\gamma_C^{-1} (\bz,\by) =
	\int \dd[3]{\bz} \gamma_C (\bx-\bz)\gamma_C^{-1} (\bz-\by) = \delta (\bx-\by).
\end{equation}
This leads to the stochastic equation
\begin{equation}
	\prtq{\dd{\ket{\psi_t}}}_{\rm meas}
	=
	\prtq{\int \dd[3]{\bx} \tmuR (\bx) \dd{W_t} (\bx) - \frac{1}{2}\int \dd[3]{\bx}\dd[3]{\by} \gamma_C (\bx,\by) \tmuR(\bx)\tmuR(\by) \dd{t}}\ket{\psi_t},
\end{equation}
where $\prtq{\dd{\ket{\psi_t}}}_{\rm meas}$ denotes the infinitesimal variation due to the measurements. Notice that $\gamma_C (\bx,\by)$ enters in the stochastic equation while the measurement error is defined through $\gamma_C^{-1} (\bx,\by)$. This can be understood intuitively: the less information is extracted from the measurement ($\gamma_C^{-1} (\bx,\by)$ is somewhat \enquote{large}), the less the measured system is disturbed ($\gamma_C (\bx,\by)$ is somewhat \enquote{small}). 

The master equation can be obtained by averaging over the stochastic equation for the density matrix: $\dd{\rho_t} = \mbE \prtq{\dd{\sigma_t}}= \mbE \prtq{\dyad{\dd{\psi_t}}{\psi_t} + \dyad{\psi_t}{\dd{\psi_t}} + \dyad{\dd{\psi_t}}}$. One gets
\begin{equation}\label{APPeq:TDModelsMasterEquationMeasurementPart}
	\dv{t} \rho_t = - \frac{1}{2}\int \dd[3]{\bx}\dd[3]{\by} \gamma_C (\bx,\by) \comm{\hmu_{r_C} (\bx)}{\comm{\hmu_{r_C} (\by)}{\rho_t}}.
\end{equation}
The CSL model is recovered by choosing $\gamma_C (\bx,\by) = (\gamma_{\rm CSL}/m_0^2) \delta (\bx-\by)$ and making the substitution $\dd{W_t} \rightarrow (\sqrt{\gamma_{\rm CSL}}/m_0) \dd{W_t}$. The DP model is instead obtained by choosing $\gamma_C (\bx,\by) = -(1/2\hbar)\mcV(\bx,\by)$, where we recall that $\mcV(\bx,\by)=-G\abs{\bx-\by}^{-1}$. 

\subsection{Adding the Gravitational Feedback\label{APPSubsec:WeaklyContinuousMonitoring_GravitationalFeedback}}

Newtonian gravity can be obtained by introducing the classical potential
\begin{equation}
	\Phi_{C} (t,\bx) = \int \dd[3]{\by} \mcV(\bx-\by)\mu_{t} (\by),
\end{equation}
where $\mcV(\bx-\by)= - G \abs{\bx-\by}^{-1}$ and $G$ is Newton's constant. We immediately see how taking the average $\mbE \prtq{\Phi_{C} (t,\bx)}$ gives the usual semiclassical Newtonian potential when considering distances much higher than $r_C$.

The procedure followed in Ref.~\cite{Tilloy2016CSLGravity} is akin to a measurement and feedback procedure~\cite{Book_Wiseman2009Measurement,Book_Jacobs2014MeasurementTheory} with a detector performing a continuous weak measurement at each spatial point. The feedback Hamiltonian chosen in Ref.~\cite{Tilloy2016CSLGravity,GaonaReyes2021GravitationalFeedback} is 
\begin{equation}
	\hH_{\rm fb} (t) 
	= \int \dd[3]{\bx} \Phi_{C} (t,\bx) \hmu_{r_G} (\bx)
	= \int \dd[3]{\bx} \dd[3]{\by} \mcV(\bx-\by) \mu_t (\by) \hmu_{r_G} (\bx),
\end{equation}
where we remark that the mass operator appearing in $H_{\rm fb} (t)$ is also smeared. Notice that, in contrast to Refs.~\cite{Tilloy2016CSLGravity,Tilloy2017LeastDecoherence,Piccione2025NewtonianPSL}, here we are not assuming that $\hmu_{r_G} = \hmu_{r_C}$

Implementing the feedback one gets the following three contributions to the differential of the wavefunction:
\begin{equations}\label{eq:TDFeedbackSchrodingerEquation}
	\prtq{\dd{\ket{\psi_t}}}_{\rm meas}
	&=
	\prtq{\int \dd[3]{\bx} \tmuR (\bx) \dd{W_t} (\bx) - \frac{1}{2}\int \dd[3]{\bx}\dd[3]{\by} \gamma_C (\bx,\by) \tmuR(\bx)\tmuR(\by) \dd{t}}\ket{\psi_t},\\
	%%%%%%%%%%%%%%%%%%%%%%%%%%%%
	\prtq{\dd{\ket{\psi_t}}}_{\rm fb} &= \prtqB{
		-\frac{i}{\hbar} \int \dd[3]{\bx}\dd[3]{\by} \mcV (\bx-\by) \ev{\hmu_{r_C} (\by)} \hmu_{r_G} (\bx) \dd{t}
		- \frac{i}{2\hbar}\int \dd[3]{\bx}\dd[3]{\by} (\mcV \circ \gamma_C^{-1}) (\bx-\by) \hmu_{r_G} (\bx) \dd{W_t} (\by) +\\
		&\qquad 
		- \frac{1}{8 \hbar^2} \int \dd[3]{\bx}\dd[3]{\by} (\mcV \circ \gamma_C^{-1} \circ \mcV) (\bx-\by) \hmu_{r_G} (\bx) \hmu_{r_G} (\by)\dd{t}}\ket{\psi_t},\\
	%%%%%%%%%%%%%%%%%%%%%%%%%%%%%%%%%%%%%%%%%%%%%%%%%%%%%%%%%%%%%%%%%%
	\prtq{\dd{\ket{\psi_t}}}_{\rm corr} &=  \prtq{-\frac{i}{2\hbar} \int \dd[3]{\bx}\dd[3]{\by} \mcV(\bx-\by) \tmuR (\by) \hmu_{r_G} (\bx)\dd{t}}\ket{\psi_t}.
\end{equations}
From the above equations, the master equation can be obtained by averaging over the stochastic master equation. The result is
\begin{multline}\label{APPeq:TD_GravitationalMasterEquation}
\dv{t} \rho_t =  
-\frac{i}{\hbar}\comm{\int \dd[3]{\bx}\dd[3]{\by} \frac{\mcV (\bx-\by)}{2} \hmu_{r_C} (\bx) \hmu_{r_G} (\by)}{\rho_t}
+\\
- \frac{1}{2} \int \dd[3]{\bx}\dd[3]{\by} \gamma_C (\bx,\by) \comm{\hmu_{r_C} (\bx)}{\comm{\hmu_{r_C} (\by)}{\rho_t}}
- \frac{1}{2} \int \dd[3]{\bx}\dd[3]{\by} \gamma_G (\bx,\by) \comm{\hmu_{r_G} (\bx)}{\comm{\hmu_{r_G} (\by)}{\rho_t}}.
\end{multline}
where we defined $\gamma_G (\bx,\by) := (2\hbar)^{-2}(\mcV \circ \gamma_C^{-1} \circ \mcV) (\bx,\by)$. In Fourier space, $\tl{\gamma}_G (\bk) = \frac{G^2}{2\pi \hbar^2 \bk^4} \frac{1}{\tl{\gamma}_C (\bk)}$ because $\tl{V}(\bk) = -\sqrt{2/\pi}G/\bk^2$. If we had $\hmu_{r_C} = \hmu_{r_G}$, then we would have
\begin{equation}
	\dv{t} \rho_t =  
	-\frac{i}{\hbar}\comm{\int \dd[3]{\bx}\dd[3]{\by} \frac{\mcV (\bx-\by)}{2} \hmu_{r_C} (\bx) \hmu_{r_C} (\by)}{\rho_t} 
	- \frac{1}{2} \int \dd[3]{\bx}\dd[3]{\by} \mcD (\bx,\by) \comm{\hmu_{r_C} (\bx)}{\comm{\hmu_{r_C} (\by)}{\rho_t}},
\end{equation}
where
\begin{equation}
	\mcD (\bx,\by) = \gamma_C (\bx,\by) + \frac{1}{4 \hbar^2}(\mcV \circ \gamma_C^{-1} \circ \mcV) (\bx,\by).
\end{equation}
Comparing Eq.~\eqref{APPeq:TD_GravitationalMasterEquation} with Eq.~\eqref{APPeq:TDModelsMasterEquationMeasurementPart} one sees that a Hamiltonian term appears due to the feedback. This term is exactly equal to the standard quantization of the Newtonian potential when $r_C \rightarrow 0$ and $r_G \rightarrow 0$. Thus, in the continuous weak monitoring approach, this term is responsible for reproducing Newtonian gravitation but it also predicts modifications of it at lengthscales lower or similar to $r_C$ and $r_G$. 

\subsection{Decoherence Rate for Isolated Particles and Rigid Bodies\label{APPSubsec:WeaklyContinuousMonitoring_DecoherenceRate}}

Let us start by considering a single particle of mass $m$.
From Eq.~\eqref{APPeq:TD_GravitationalMasterEquation}, the decoherence rate is given by
\begin{equation}\label{APPeq:TD_SingleParticleDecoherenceRate}
\Gamma(\bd) 
= 
m^2 (2\pi)^{3/2} \int \dd[3]{\bk} 
\prtq{\tl{\gamma}_C (k) \abs{\tl{g}_{r_C} (k)}^2
+ \frac{\tl{\mcV}^2(k)}{4 \hbar^2 \tl{\gamma}_C (k)}\abs{\tl{g}_{r_G} (k)}^2}
\prtq{1-\cos(\bk \cdot \bd)},
\end{equation}
where $\tl{g}_{r_C} (k)$ and $\tl{\mcV} (k)$ are the diagonals of the respective Fourier transforms, which are diagonal because both $\gamma_c$ and $\mcV$ are translation invariant. Here, we can already see why the calculations in Ref.~\cite{Tilloy2017LeastDecoherence} necessitate that $g_{r_C}=g_{r_G}$: the ``Principle of Least Decoherence'' was applied by minimizing the decoherence rate at the level of each single mode, which would give
\begin{equation}
\tl{\gamma}_C (k) = -\frac{\tl{\mcV} (k)}{2\hbar} \frac{\abs{\tl{g}_{r_G} (k)}}{\abs{\tl{g}_{r_C} (k)}}.
\end{equation}
This gives the DP model only if $\tl{g}_{r_C} (k) = \tl{g}_{r_G} (k)$.

Going on, by performing the spherical average of Eq.~\eqref{APPeq:TD_SingleParticleDecoherenceRate}, one gets
\begin{equation}
\Gamma(d)
=
4\pi m^2 (2\pi)^{3/2}\int_0^{\infty} \dd{k} k^2
\prtq{\tl{\gamma}_C (k) \abs{\tl{g}_{r_C} (k)}^2
+ \frac{\tl{\mcV}^2(k)}{4 \hbar^2 \tl{\gamma}_C (k)}\abs{\tl{g}_{r_G} (k)}^2}\prtq{1-j_0 (d k)},
\qquad
j_0 (x):= \frac{\sin(x)}{x}.
\end{equation}
Then, assuming reasonable smearing functions, we get that
\begin{equation}
\Gamma(d \gg r_C,r_G)
\simeq
4\pi m^2 (2\pi)^{3/2}\int_0^{\infty} \dd{k} k^2
\prtq{\tl{\gamma}_C (k) \abs{\tl{g}_{r_C} (k)}^2
+ \frac{\tl{\mcV}^2(k)}{4 \hbar^2 \tl{\gamma}_C (k)}\abs{\tl{g}_{r_G} (k)}^2},
\end{equation}
which, given the lengths $r_C$ and $r_G$, still depends on the choices made for the two smearing distributions.

Let us now consider instead the case of a rigid body. In particular, how to estimate the decoherence rate of its CoM in a spatial superposition. However, we keep fixed the orientation of the rigid body. In (for example) Ref.~\cite{Piccione2025ExploringMassDependence}, it is seen how one can approximate the action of the smeared mass density operator on the CoM:
\begin{equation}\label{APPeq:CoM_MassDensityOperatorV1}
\hmu_{r_C} (\bx) \rightarrow 
\mu_{\rm CM} (\bx-\hbQ) \simeq \sum_k m_k g_{r_C} \prt{\bx-\hbQ-\tl{\bq}_k (r_0)},
\qquad
\hbQ := \frac{1}{M}\sum_{k} m_k \hbq_k,
\end{equation}
where $r_0$ denotes the equilibrium internal coordinates of the rigid body and $\tl{\bq}_k (r_0)$ the position of the $k$-th particle with respect to the CoM. As long as there are many atoms within a radius $r_C$ and the mass density of the body varies over scales much larger than $r_C$, for the purposes of estimating the decoherence rate of a macroscopic body one can basically ignore the smearing in the sense that every reasonable smearing should give the same result, i.e., the mass density classically $\varrho(\bx)$ associated to that rigid body.
In other words, body density varies on scales much larger than $r_C$, with many atoms per $r_C$-volume. Then, assuming that the same applies to the smearing with $g_{r_G}$, from Eq.~\eqref{APPeq:TD_GravitationalMasterEquation} one gets (ignoring the Hamiltonian)
\begin{equation}
\dot{\rho}^{\rm CM}_t (\bX,\bY)
=
-\Gamma(\bX-\bY)\rho^{\rm CM}_t (\bX,\bY),
\qquad
\Gamma(\bD) = \int \dd[3]{\bz} \varrho(\bz) \int \dd[3]{\bz'} \mcD(\bz-\bz')\prtq{\varrho(\bz')-\varrho(\bz'+\bD)}.
\end{equation}
The above result is, unlike the single particle case, independent of the smearing distribution. 
One can also write the above quantity in Fourier space:
\begin{equation}
\Gamma(\bD) = (2\pi)^{3/2} \int \dd[3]{\bk} \tl{\mcD}(\bk)\abs{\tl{\varrho}(\bk)}^2 \prtq{1-\cos(\bk\cdot\bD)},
\end{equation}
from which follows that, applied to the center of mass of large macroscopic bodies, the ``Principle of Least Decoherence'' would lead to the DP correlator independently of which smearing distributions are used.

\clearpage
\section{Computation of formulas for obtaining the strongest experimental bounds\label{APPSec:ConnectionExperimentalBounds}}

\subsection{Stochastic Unraveling of Continuous weak Monitoring Models Master Equations\label{APPSubsec:UnitaryUnravelingTDModels}}

It is usually convenient to unravel the master equation of a spontaneous collapse model in terms of a stochastic potential. This is a standard procedure since Ref.~\cite{Fu1997SpontaneousRadiation}. Here, we generalize this procedure to the case of Eqs.~\eqref{eq:HybridModelsEquation} and~\eqref{eq:HybridModelsEquation_Components} of the main text.

Let us consider a Markovian master equation of the following kind:
\begin{equation}
\dot{\rho}_t = -\frac{i}{\hbar}\comm{\hH}{\rho_t} + \mcL_C [\rho_t] + \mcL_G [\rho_t],
\qquad
\mcL_A [\rho_t] = - \frac{1}{2} \int \dd[3]{\bx}\dd[3]{\by} \gamma_A (\bx,\by) \comm{\hmu_{r_A} (\bx)}{\comm{\hmu_{r_A} (\by)}{\rho_t}},
\quad
A=C,G.
\end{equation}
We want to prove that there exists a unitary unraveling given by the following stochastic potentials:
\begin{equation}
\hW_A (t) = \hbar \int \dd[3]{\bx} \hmu_{r_A} (\bx) w_A (t,\bx),
\end{equation}
where $w_A (t,\bx)$ are white-in-time noises characterized by 
\begin{equation}
\mbE\prtq{w_A (t,\bx)}=0,
\qquad
\mbE \prtq{w_A (t,\bx)w_A (s,\by)}= \delta\prt{t-s} \gamma_A (\bx,\by),
\qquad
\mbE \prtq{w_C (t,\bx) w_G (s,\by)}= 0.
\end{equation}

To prove the validity of the unraveling one starts with a density matrix $\rho_t$. Since both the master equation and the alleged unraveling are Markovian, at any time $t$ we can take $\rho_t$ to be the quantum state already averaged over the noise up to that time. For every realization of the noise, we can evolve it in time to $t+\delta t$, up to second order in $\delta t$:
\begin{equation}
	U(t+\delta t,t) = \mcT \exp{-\frac{i}{\hbar}\int_t^{t+\delta t} \hH'(s) \dd{s}}
	\simeq 1 - \frac{i}{\hbar}\int_t^{t+\delta t} \hH'(s) \dd{s} -\frac{1}{\hbar^2}\int_t^{t+\delta t}\int_t^{t_1}  \hH'(t_1) \hH'(t_2) \dd{t_1}\dd{t_2},
\end{equation}
where $\hH'(t) = \hH + \hW_C (t) + \hW_G (t)$.
The average density matrix at time $t+\delta t$ is obtained by taking the average over the noise of $\rho_{t+\delta t} = U(t+\delta t,t)\rho_t U^\dg(t+\delta t,t)$ and keeping only terms up to first order in $\delta t$. 
The heuristic argument\footnote{This is connected to the idea of writing $w(t,\bx) = \dd{W_t (\bx)}/\dd{t}$, where $\dd{W_t} (\bx)$ is the generalized Wiener increment.} to do so is to consider $\prtq{w(t,\bx)}\sim \delta t^{-1/2}$ so that, for example, a term like $\int \hH \hW_A (t_2)\dd{t_1}\dd{t_2}$ and $\int \hW_A (t_1)\hW_A(t_2)\hW_A(t_3) \dd{t_1}\dd{t_2}\dd{t_3}$ are both of higher order and have to be neglected. Then, one must use equalities such as $\mbE \prtq{\int_t^{t+\delta t} \hW_A(s)\dd{s}}=0$,
\begin{equation}
	\mbE\prtq{\int_t^{t+\delta t}\int_t^{t+\delta t} \dd{t_1} \dd{t_2} \hW_A(t_1)\rho_t \hW_A (t_2)}
	= \hbar^2 \delta t \int \dd[3]{\bx}\dd[3]{\by} \gamma_A(\bx,\by) \hmu_{r_A}(\bx)\rho_t \hmu_{r_A}(\by),
\end{equation}
and
\begin{multline}
	\mbE\prtq{\int_t^{t+\delta t}\int_t^{t_1} \dd{t_1} \dd{t_2} \hW_A(t_1)\hW_A(t_2)\rho_t}
	= \hbar^2 \prtq{\int_t^{t+\delta t}\int_t^{t_1} \delta(t_1-t_2)\dd{t_1}\dd{t_2}} \int \dd[3]{\bx}\dd[3]{\by} \gamma_A(\bx,\by)\hmu_{r_A}(\bx)\hmu_{r_A} (\by)\rho_t
	=\\=
	\frac{\delta t}{2}\hbar^2 \int \dd[3]{\bx}\dd[3]{\by} \gamma_A(\bx,\by)\hmu_{r_A}(\bx)\hmu_{r_A} (\by)\rho_t.
\end{multline}
The master equation is finally obtained by writing $\dot{\rho}_t = \lim_{\delta t \rightarrow 0} (\rho_{t+\delta t} - \rho_t)/\delta t$.

\subsection{Radiation of free charged particles}

Here, we compute the emission rate of a free particle by means of a semiclassical derivation based on Larmor's formula. This approach can be found in Refs.~\cite{Adler2007Bounds,Donadi2014RadiationEmission,Donadi2021NovelCSLBounds}. This subsection closely follows the derivation done in Appendix C of Ref.~\cite{Piccione2025ExploringMassDependence}.

The power of radiation emitted by a point particle with charge $q$ is
\begin{equations}\label{APPeq:LarmorFormula}
P(t) 
&= \frac{q^2}{6 \pi \varepsilon_0 c^3} a^2 (t)
= \frac{q^2}{6 \pi \varepsilon_0 c^3} \sum_{j=1}^{3} \frac{1}{2\pi} \int \dd{\omega}\dd{\nu} e^{-i(\nu+\omega)t} \tl{a}_j (\nu) \tl{a}_j (\omega),
\qquad
\tl{a}_j (\omega) := \int \frac{\dd{t}}{\sqrt{2\pi}} a_j (t) e^{i \omega t},
\\
P(t) &= \int_0^{\infty} \dd{\omega} \hbar \omega \dv{\Gamma (t)}{\omega}, 
\end{equations}
where the first equation in the first line is Larmor's formula and $\tl{a}_j (\omega)$ is the Fourier transform of $a_j (t)$. The equation in the second line is the integration over all frequencies of the rate of photons emitted at a given frequency times their energy. 
According to our unitary stochastic unraveling (see Appendix~\ref{APPSubsec:UnitaryUnravelingTDModels}), the free particle is subjected to the following acceleration:
\begin{multline}
a_j(t) 
= 
\frac{i}{\hbar m}\comm{\hW_C (t) + \hW_G (t)}{\hp_j}
= 
i \int \dd[3]{\bx} \comm{g_{r_C}(\bx-\hbq)w_C (t,\bx) + g_{r_G}(\bx-\hbq)w_G (t,\bx)}{\hp_j}
=\\
= 
\frac{\hbar}{\sqrt{2\pi}} \int \dd{\omega} \dd[3]{\bx} \prtg{\prtq{\partial_j g_{r_C}(\bx-\hbq)} \tl{w}_C (\omega,\bx) + \prtq{\partial_j g_{r_G}(\bx-\hbq)} \tl{w}_G (\omega,\bx)}e^{-i \omega t}.
\end{multline}
where the subscript $j$ denotes the spatial direction.
Thus, we have that
\begin{equation}
\tl{a}_j (\omega) 
= 
\hbar \int \dd[3]{\bx} \prtg{\prtq{\partial_j g_{r_C}(\bx-\hbq)} \tl{w}_C (\omega,\bx) + \prtq{\partial_j g_{r_G}(\bx-\hbq)} \tl{w}_G (\omega,\bx)}.
\end{equation}

We can now compute the average power in frequency. Using
\begin{equation}
\mbE\prtq{\tl{w}_A (\omega,\bx)}=0,
\qquad
\mbE \prtq{\tl{w}_A (\omega,\bx)\tl{w}_A (\omega',\by)}= \delta\prt{\omega+\omega'} \gamma_A (\bx,\by),
\qquad
\mbE \prtq{\tl{w}_C (\omega,\bx) \tl{w}_G (\omega',\by)}= 0,
\end{equation}
we get
\begin{multline}
\mbE[P(t)]
=
\frac{q^2}{6 \pi \varepsilon_0 c^3} \sum_{j=1}^{3} \frac{1}{2\pi} \int \dd{\omega}\dd{\nu} e^{-i(\nu+\omega)t} \mbE[\ev{\tl{a}_j (\nu) \tl{a}_j (\omega)}]
=\\
=
\frac{q^2 \hbar^2}{6 \pi \varepsilon_0 c^3} \sum_{j=1}^{3} \int \dd[3]{\bx}\dd[3]{\by} \int \frac{\dd{\omega}\dd{\nu}}{2\pi} e^{-i(\nu+\omega)t}\delta(\nu+\omega)
\times \\ \times
\prtg{
\prtq{\partial_j g_{r_C} (\bx-\hbq)}\prtq{\partial_j g_{r_C} (\by-\hbq)}\gamma_C (\bx,\by)
+
\prtq{\partial_j g_{r_G} (\bx-\hbq)}\prtq{\partial_j g_{r_G} (\by-\hbq)}\gamma_G (\bx,\by)
}
=\\
=
\frac{q^2 \hbar^2}{12 \pi^2 \varepsilon_0 c^3}
\int \dd{\omega}
\int \dd[3]{\bx}\dd[3]{\by}
\prtg{\prtq{\nabla g_{r_C} (\bx-\hbq)}\cdot \prtq{\nabla g_{r_C} (\by-\hbq)}\gamma_C (\bx,\by)
+
\prtq{\nabla g_{r_G} (\bx-\hbq)}\cdot \prtq{\nabla g_{r_G} (\by-\hbq)}\gamma_G (\bx,\by)}.
\end{multline}
So, by comparison with Eq.~\eqref{APPeq:LarmorFormula} we get that\footnote{Notice that, in the second line of Eq.~\eqref{APPeq:LarmorFormula}, the integration goes from $0$ to $+\infty$.}
\begin{equation}
\mbE\prtq{\dv{\Gamma (t)}{\omega}}
\!=\! 
\frac{q^2 \hbar}{6 \pi^2 \varepsilon_0 c^3 \omega}
\!
\int\! \dd[3]{\bx}\dd[3]{\by}
\prtg{\prtq{\nabla g_{r_C} (\bx-\hbq)}\cdot \prtq{\nabla g_{r_C} (\by-\hbq)}\gamma_C (\bx,\by)
+
\prtq{\nabla g_{r_G} (\bx-\hbq)}\cdot \prtq{\nabla g_{r_G} (\by-\hbq)}\gamma_G (\bx,\by)}.
\end{equation}
Notice how both the total power and the emission rate at a given frequency are proportional to $I_{\gamma_C}[g_{r_C}] + I_{\gamma_G}[g_{r_G}]$ (see Eq.~\eqref{eq:GenericHeatingRateFunctional}). Indeed, one can write
\begin{equation}
\mbE[P(t)]
=
\frac{q^2 \hbar^2}{6 \pi^2 \varepsilon_0 c^3}\prt{I_{\gamma_C}[g_{r_C}] + I_{\gamma_G}[g_{r_G}]}\int \dd{\omega} ,
\qquad
\mbE\prtq{\dv{\Gamma (t)}{\omega}}
=
\frac{1}{\omega}\times
\frac{q^2 \hbar}{3 \pi^2 \varepsilon_0 c^3} \prt{I_{\gamma_C}[g_{r_C}] + I_{\gamma_G}[g_{r_G}]}.
\end{equation}

\subsection{Radiation of a large rigid body}

Let us consider a system governed by Eq.~\eqref{eq:HybridModelsEquation}.
Exploiting the unitary unraveling of Sec.~\ref{APPSubsec:UnitaryUnravelingTDModels} one can see how each particle, say the $j$-th one, is subjected to the force
\begin{equation}
\hbF_j = 
\hbF_j^{\rm (QM)} + \hbF_j^{\rm (Grav.)}+
\hbF_j^{(C)} + \hbF_j^{(G)},
\end{equation}
where
\begin{equation}
\hbF_j^{\rm (QM)} := \frac{i}{\hbar}\comm{\hH}{\hbp_j},
\qquad
\hbF_j^{\rm (Grav.)} := \frac{i}{\hbar}\comm{\hV_{r_C,r_G}}{\hbp_j},
\qquad
\hbF_j^{(A)} := \frac{i}{\hbar}\comm{\hW_A (t)}{\hbp_j},
\end{equation}
where we recall that $A=C,G$. Charged particles subject to acceleration emit radiation~\cite{Fu1997SpontaneousRadiation,Donadi2014RadiationEmission,Donadi2015Radiation}, an effect which has been used to set experimental bounds on collapse models parameters~\cite{Donadi2021UndergroundTest,Donadi2021NovelCSLBounds,MAJORANACollaboration2022WaveFunctionCollapse,Piccione2025ExploringMassDependence,Piccione2026EnergyIncreaseGPSL,XENONCollaboration2026CollapseConstraints}. The usual way of computing this effect is by means of a semiclassical approach in which the above forces become classical stochastic forces acting on point particles. Then, via classical electrodynamics, one can deduce the expected emission due to additional noise introduced by spontaneous collapse theories and hybrid classical-quantum theory of Newtonian gravity\footnote{Strictly speaking, due to the smearing, $\hbF_j^{\rm (Grav.)}$ is not exactly equal to the force one would obtain from the exact Newtonian potential. However, the gravitational contribution to the radiation is already negligible by itself in typical scenarios, the deviation due to the smearing is expected to be even smaller. Therefore, we only consider the emission due to $\hbF_j^{(A)}$ with $A=C,G$.}. The explicit expressions of the stochastic forces are
\begin{equation}
\hbF_j^{(A)}
=
\hbar m_j \int \dd[3]{\bx} w_A (t,\bx) \nabla g_{r_A}(\bx-\hbq).
\end{equation}

As in the appendix D.2 of Ref.~\cite{Piccione2025ExploringMassDependence}, our starting point comes from the Appendix of Ref.~\cite{Donadi2021NovelCSLBounds}. This gives us approximate formulas for the power of the emitted radiation at long distances from the sources. This formula is independent of the spontaneous collapse model we use. We have that\footnote{Notice that we adapted the formula to our convention on Fourier transforms.}
\begin{equation}
P(t) = \frac{1}{32 \pi^3 \varepsilon_0 c^3} \intmp \dd{\omega} \intmp \dd{\nu} e^{i(\omega+\nu)(t-R_{sp}/c)} \sum_{i,j} q_i q_j J_{i,j} (\omega,\nu),
\end{equation}
where $R_{sp}$ is the radius of the sphere over which we measure the emitted radiation, $\bn$ is the unit normal vector on the sphere surface and pointing outward, and
\begin{equation}
J_{i,j} (\omega,\nu) =
4 \pi \prtq{ \ddot{\br}_i (\omega) \cdot \ddot{\br}_j (\nu) \frac{(b^2-1)\sin(b) + b\cos(b)}{b^3} - \ddot{\br}^z_i (\omega) \ddot{\br}^z_j (\nu) \frac{(b^2-3)\sin(b) + 3b\cos(b)}{b^3}},
\end{equation}
where $b = \abs{\omega \ev{\br_i} + \nu \ev{\br_j}}/c$, with $\br_j$ the position of the $j$-th particle, and, in this subsection, we are using the convention that upper indices denote the spatial direction.

In frequency domain, the acceleration felt by each particle is given by
\begin{equation}
\ddot{\br}_k^j (\omega)
=
\frac{\hF_k^{j,(C)} + \hF_k^{j,(G)}}{m_k}
=
\hbar \int \dd[3]{\bx} \prtq{\tl{w}_C (\omega,\bx)\partial^j g_{r_C} (\bx-\hbq_k)+\tl{w}_G (\omega,\bx)\partial^j g_{r_G} (\bx-\hbq_k)}.
\end{equation}
Then, we have that
\begin{equations}
\mbE\prtq{\ev{\ddot{\br}_i (\omega) \cdot \ddot{\br}_j (\nu)}}
&=
\hbar^2 \delta(\omega+\nu) \sum_A \int \dd[3]{\bx}\dd[3]{\by}
\gamma_A (\bx,\by) 
\ev{\prtq{\nabla g_{r_A} (\bx-\hbq_i)}\cdot \prtq{\nabla g_{r_A} (\by-\hbq_j)}},
\\
\mbE\prtq{\ev{\ddot{\br}_i^z (\omega) \cdot \ddot{\br}_j^z (\nu)}}
&=
\hbar^2 \delta(\omega+\nu) \sum_A \int \dd[3]{\bx}\dd[3]{\by}
\gamma_A (\bx,\by) 
\ev{\prtq{\partial^z g_{r_A} (\bx-\hbq_i)}\prtq{\partial^z g_{r_A} (\by-\hbq_j)}}.
\end{equations}
We can now make two approximations. First, the position of each particle within each body can vary only on scales much smaller than $r_A$ so that we can bring the expectation value inside the smearing functions. Second, we can assume that, 
for the vast majority of particles within the emitting body, all spatial directions are effectively equivalent so that we can write $\mbE[\ddot{\br}_{k}^j (\omega) \cdot \ddot{\br}_{k'}^j (\nu)] = (1/3) \mbE[\ddot{\br}_{k} (\omega) \cdot \ddot{\br}_{k'} (\nu)]$, for all directions $j$. Thus, we get that
\begin{equation}
\mbE[J_{i,j} (\omega,\nu)] 
= \frac{8 \pi \sin(b)}{3 b}\mbE[\ddot{\br}_{i} (\omega) \cdot \ddot{\br}_{j} (\nu)]
= \frac{8 \pi \hbar^2}{3}\delta (\omega+\nu)\ \mathrm{sinc}\prt{\frac{2\pi\abs{\ev{\br_i}-\ev{\br_j}}}{\lambda_\omega}} f_{i,j},
\qquad
\mathrm{sinc}(x) = \frac{\sin(x)}{x},
\end{equation}
where $\lambda_\omega = 2\pi c/\omega$ is the wavelength of the emitted radiation and
\begin{equation}
f_{i,j} := \sum_A \int \dd[3]{\bx}\dd[3]{\by}
\gamma_A (\bx,\by) 
\ev{\prtq{\nabla g_{r_A} (\bx-\hbq_i)}\cdot \prtq{\nabla g_{r_A} (\by-\hbq_j)}}.
\end{equation}
Substituting this in the expression for the power, we get
\begin{equation}
\mbE[P(t)] 
= 
\frac{\hbar^2}{12 \pi^2 \varepsilon_0 c^3}\sum_{i,j} q_i q_j f_{i,j} \int \dd{\omega} \mathrm{sinc}\prt{\frac{2\pi\abs{\ev{\br_i}-\ev{\br_j}}}{\lambda_\omega}}.
\end{equation}
Comparing this result with the decomposition of $P(t)$ as an integral over the emission rate one gets
\begin{equation}
\mbE \prtq{\dv{\Gamma (t)}{\omega}} = \frac{1}{\omega}\times
\frac{\hbar}{6 \pi^2 \varepsilon_0 c^3} \sum_{i,j} q_i q_j f_{i,j}\ \mathrm{sinc}\prt{\frac{2\pi\abs{\ev{\br_i}-\ev{\br_j}}}{\lambda_\omega}}.
\end{equation}

Now, as done in Refs.~\cite{Donadi2021NovelCSLBounds,Donadi2021UndergroundTest,Piccione2025ExploringMassDependence}, we consider that most radiation comes from considering these two relevant regimes\footnote{A more refined analysis is done in Ref.~\cite{Piscicchia2024SpontaneousEmissionRefined}, where wavelengths comparable with $\abs{\ev{\br_i}-\ev{\br_j}}$ are also considered.}:
\begin{equations}
\abs{\ev{\br_i}-\ev{\br_j}} \gg \lambda_\omega
&\implies
q_i q_j f_{i,j}\mathrm{sinc}\prt{\frac{2\pi\abs{\ev{\br_i}-\ev{\br_j}}}{\lambda_\omega}}
\rightarrow
0,
\\
\abs{\ev{\br_i}-\ev{\br_j}} \ll \lambda_\omega, r_C, r_G
&\implies
q_i q_j f_{i,j}\mathrm{sinc}\prt{\frac{2\pi\abs{\ev{\br_i}-\ev{\br_j}}}{\lambda_\omega}}
\rightarrow
q_i q_j f_{j,j}.
\end{equations}
However, we have that $f_{j,j} = 2\prt{I_{\gamma_C}[g_{r_C}] + I_{\gamma_G}[g_{r_G}]}$ so that, to a good approximation:
\begin{equation}
\mbE \prtq{\dv{\Gamma (t)}{\omega}} 
\propto
I_{\gamma_C}[g_{r_C}] + I_{\gamma_G}[g_{r_G}].
\end{equation}
Which means that, even for a large rigid body, minimizing the heating rate also greatly reduces the amount of radiation emitted. In particular, under the approximations we used, minimizing the radiation of a large rigid body is equivalent to minimizing the general heating rate.

\subsection{Diffusion of the CoM of a rigid body}

Here, we compute the diffusion of the CoM of an isolated rigid body. First, we define the position and momentum operators associated to it:
\begin{equation}
\hbQ := \frac{1}{M}\sum_{k} m_k \hbq_k,
\quad
\hbP := \sum_k {\hbp}_k,
\qquad
M = \sum_k m_k,
\end{equation}
so that the CoM diffusion is quantified by the kinetic energy of the CoM, i.e., $\hat K := \hbP^2/2M$. By assumption, the CoM momentum operator commutes with the Hamiltonian of the system because we assume that the rigid body is isolated. The heating rate of the CoM is then [cf. Eqs.~\eqref{eq:HybridModelsEquation} and~\eqref{eq:HybridModelsEquation_Components}]
\begin{equation}
\dot{E}^{\rm CoM}_t
=
\dv{t} \Tr{\hat K \rho_t}
=
\frac{\hbar^2}{2 M} \sum_{A=C,G} \int \dd[3]{\bx}\dd[3]{\by} \gamma_A (\bx-\by) \Tr \prtg{\prt{ \nabla \hmu_{r_A} (\bx) \cdot \nabla \hmu_{r_A} (\by)} \rho_t}.
\end{equation}
Using Eq.~\eqref{APPeq:CoM_MassDensityOperatorV1} and employing the so-called macroscopic density approximation~\cite{Review_Bassi2003Dynamical}\footnote{See Secs. 8.3 and 8.4 of Ref.~\cite{Review_Bassi2003Dynamical}.} we can write
\begin{equation}
\hmu_{r_A} (\bx)
\simeq
\int \dd[3]{\bz} \mu(\bz) g_{r_A} \prt{\bx-\hbQ-\bz},
\qquad
\nabla \hmu_{r_A} (\bx)
\simeq
\int \dd[3]{\bz} \mu(\bz) \nabla g_{r_A} \prt{\bx-\hbQ-\bz},
\end{equation}
where $\mu(\bz)$ is the classical macroscopic mass density associated to the rigid body\footnote{ 
This classical macroscopic density is given for an arbitrary but fixed orientation of the rigid body. Moreover, for simplicity, this macroscopic density should be considered centered, i.e., $\int \dd[3]{\bx} \bx \mu(\bx)=0$. In other words, $\mu(\bx)$ is the classical macroscopic density the rigid body would have if its CoM was at the coordinate origin.}.
Thus, we get
\begin{equation}
\dot{E}^{\rm CoM}_t
\simeq
\frac{\hbar^2}{2 M} \sum_{A=C,G} 
\int \dd[3]{\bx}\dd[3]{\by}\dd[3]{\bz}\dd[3]{\bz'}\dd[3]{\bQ} p^{\rm (CoM)}_t (\bQ) \gamma_A (\bx-\by) \mu (\bz)\mu (\bz') \prtq{\nabla g_{r_A} \prt{\bx-\bQ-\bz}}\cdot\prtq{\nabla g_{r_A} \prt{\by-\bQ-\bz'}},
\end{equation}
where $p^{\rm (CoM)}_t (\bQ)$ is the probability density distribution associated to the CoM position. Introducing $\ba= \bx-\bQ-\bz$ and $\bb = \by-\bQ-\bz'$ we see that
\begin{multline}
\dot{E}^{\rm CoM}_t
\simeq
\frac{\hbar^2}{2 M} \sum_{A=C,G} 
\int \dd[3]{\ba}\dd[3]{\bb}\dd[3]{\bz}\dd[3]{\bz'}\dd[3]{\bQ} p^{\rm (CM)}_t (\bQ) \gamma_A (\ba-\bb + \bz-\bz') \mu (\bz)\mu (\bz') \prtq{\nabla g_{r_A} (\ba)}\cdot\prtq{\nabla g_{r_A} (\bb)}
=\\
=
\frac{\hbar^2}{2 M} \sum_{A=C,G} 
\int \dd[3]{\ba}\dd[3]{\bb}\dd[3]{\bz}\dd[3]{\bz'} \gamma_A (\ba-\bb + \bz-\bz') \mu (\bz)\mu (\bz') \prtq{\nabla g_{r_A} (\ba)}\cdot\prtq{\nabla g_{r_A} (\bb)}
=\\
=
\frac{\hbar^2}{2 M} \sum_{A=C,G} 
\int \dd[3]{\ba}\dd[3]{\bb} \gamma^{\rm (CoM)}_A (\ba-\bb) \prtq{\nabla g_{r_A} (\ba)}\cdot\prtq{\nabla g_{r_A} (\bb)},
\end{multline}
where $\gamma^{\rm (CoM)}_A (\br) := \int \dd[3]{\bz}\dd[3]{\bz'} \gamma_A (\br + \bz-\bz') \mu (\bz)\mu (\bz')$.
Notice that, under the macroscopic rigid body assumptions, the heating rate of the CoM is state-independent and its formula is quite similar to that of Eq.~\eqref{eq:GenericHeatingRateFunctional}.
One can also rewrite the heating rate expression as
\begin{multline}\label{APPeq:CenterOfMassHeatingRateFourierSpace}
\dot{E}^{\rm CoM}_t
\simeq
\frac{\hbar^2}{2 M} \sum_{A=C,G} 
\int \dd[3]{\bx}\dd[3]{\by}\dd[3]{\bz}\dd[3]{\bz'} \gamma_A (\bx-\by) \mu (\bz)\mu (\bz') \prtq{\nabla g_{r_A} (\bx-\bz)}\cdot\prtq{\nabla g_{r_A} (\by-\bz')}
=\\
=
\frac{\hbar^2 (2\pi)^{9/2}}{2 M} \sum_{A=C,G} 
\int 
\dd[3]{\bk}\dd[3]{\bk_1}\dd[3]{\bk_2}\dd[3]{\bk_3}\dd[3]{\bk_4}
\frac{\dd[3]{\bx}\dd[3]{\by}\dd[3]{\bz}\dd[3]{\bz'}}{(2\pi)^{12}}
(\bk_3\cdot\bk_4)\tl{\gamma}_A (\bk)\tl{\mu}^*(\bk_1)\tl{\mu}(\bk_2)\tl{g}^*_{r_A}(\bk_3)\tl{g}_{r_A}(\bk_4)
\times \\ \times
\exp{i\prtq{-\bx\cdot(\bk-\bk_3) + \by\cdot(\bk-\bk_4)+\bz\cdot(\bk_1-\bk_3)-\bz'\cdot(\bk_2-\bk_4)}}
=\\
=
\frac{\hbar^2 (2\pi)^{9/2}}{2 M} \sum_{A=C,G} 
\int \dd[3]{\bk} \bk^2 \tl{\gamma}_A (\bk) \abs{\tl{\mu}(\bk)}^2 \abs{\tl{g}_{r_A}(\bk)}^2
=
\frac{\hbar^2 (2\pi)^{9/2}}{2 M} \sum_{A=C,G} 
\int \dd[3]{\bk} \bk^2 \tl{\gamma}_A (\bk) \abs{\tl{\mu}(\bk) \tl{g}_{r_A}(\bk)}^2.
\end{multline}
For comparison, the total heating rate of $N$ particles in Fourier space can be written as (see Eq.~\eqref{eq:GenericHeatingRateFunctional})
\begin{equation}
\dot{E}_t = \frac{\hbar^2 M (2\pi)^{3/2}}{2} \sum_{A=C,G} \int \dd[3]{\bk} \bk^2 \tl{\gamma}_A (\bk) \abs{\tl{g}_{r_A}(\bk)}^2.
\end{equation}

We can further simplify the expression for the CoM diffusion rate by assuming that $\mu(\bx)$ varies over scales much larger than $r_A$. We define the smeared classical macroscopic density:
\begin{equation}
\mu_{r_A} (\bx) := \int \dd[3]{\bz} \mu(\bz)g_{r_A} (\bx-\bz),
\implies
\dot{E}^{\rm CoM}_t
\simeq
\frac{\hbar^2}{2 M} \sum_{A=C,G} 
\int \dd[3]{\bx}\dd[3]{\by} \gamma_A (\bx-\by) \prtq{\nabla \mu_{r_A} (\bx)}\cdot \prtq{\nabla \mu_{r_A} (\by)}.
\end{equation}
Now, when $\mu(\bx)$ varies over scales much larger than $r_A$, $\mu_{r_A} (\bx) \simeq \mu (\bx)$. The notable exception to this condition lies at the boundary of the rigid body, where, typically, $\mu(\bx)$ varies over scales much smaller than $r_A$. Thus, we can separate the heating/diffusion rate in two pieces:
\begin{equation}
\dot{E}^{\rm CoM}_t
\simeq
\dot{E}^{\rm CoM}_t\vert_{V} + \dot{E}^{\rm CoM}_t\vert_{\partial V},
\end{equation}
where $V$ denotes the volume (the interior) of the rigid body and $\partial V$ its boundary (its surface). We have that
\begin{equation}
\dot{E}^{\rm CoM}_t\vert_{V}
:=
\frac{\hbar^2}{2 M} \sum_{A=C,G} \int_{V} \dd[3]{\bx}\dd[3]{\by} \gamma_A (\bx-\by) \prtq{\nabla \mu (\bx)}\cdot \prtq{\nabla \mu (\by)},
\end{equation}
which is independent of the form of the smearing distribution. If the body has constant density, this term gives a vanishing contribution. We have now to find a suitable expression for $\dot{E}^{\rm CoM}_t\vert_{\partial V}$. To do this, we assume that the mass density of the rigid body is the same over all its surface and go back to the full integral but substituting the mass density of the rigid body with the one at its surface. Denoting by $\mu_S$ this surface density, in Fourier space, this gives [see Eq.~\eqref{APPeq:CenterOfMassHeatingRateFourierSpace}]
\begin{equation}
\dot{E}^{\rm CoM}_t\vert_{\partial V}
=
\frac{\hbar^2 (2\pi)^{9/2} \mu_S^2 }{2 M} \sum_{A=C,G} 
\int \dd[3]{\bk} \bk^2 \tl{\gamma}_A (\bk) \abs{\tl{\chi}_V(\bk) \tl{g}_{r_A}(\bk)}^2,
\end{equation}
where $\chi_V (\bx)$ is the indicator function associated to the shape of the rigid body.
It does not seem possible to reduce the above quantity to an expression proportional to $I_{\gamma_C}[g_{r_C}] + I_{\gamma_G}[g_{r_G}]$, as we did for the radiation. Thus, we conclude that optimizing over the diffusion of the CoM of a large rigid body would probably lead to a different smearing profile than the one we found in the main text.

\end{document}

%% file: preamble_packages.tex
\usepackage[T1]{fontenc}
\usepackage{graphicx}% Include figure files
\graphicspath{{./Figures/}}
\usepackage{dcolumn}% Align table columns on decimal point
\usepackage{bm}% bold math
\usepackage[dvipsnames]{xcolor}
\usepackage{hyperref}% add hypertext capabilities
\hypersetup{colorlinks,linkcolor={MidnightBlue},citecolor={MidnightBlue},urlcolor={MidnightBlue}}  
\usepackage[table,xcdraw]{xcolor}
\usepackage{tabularray}

\usepackage{float}
\usepackage{wrapfig}

\usepackage{tensor}

\usepackage{physics}
\usepackage{amsfonts}
\usepackage{verbatim}
\usepackage{amsmath}
\usepackage{csquotes}
\usepackage{color}
\usepackage{soul}
\usepackage{amsthm}
\usepackage{bm}
\usepackage{graphicx}
\usepackage[normalem]{ulem}
\usepackage{mathtools}

\usepackage{yfonts}

\usepackage{verbatim}

%% file: preamble_commands.tex
\newenvironment{equations}
{\begin{equation}\begin{aligned}}
		{\end{aligned}\end{equation}\ignorespacesafterend}

\newenvironment{equations*}
{\begin{equation*}\begin{aligned}}
		{\end{aligned}\end{equation*}\ignorespacesafterend}

\newcommand{\prt}[1]{\left(#1\right)}

\newcommand{\intmp}{\int_{-\infty}^{+\infty}}

\newcommand{\dg}{\dagger}
\newcommand{\tl}{\tilde}

\newcommand{\prtq}[1]{\left[#1\right]}

\newcommand{\prtg}[1]{\left\{#1\right\}}

\newcommand{\prtqB}[1]{\Bigg[#1\Bigg]}

\newcommand{\mcD}{\mathcal{D}}

\newcommand{\mcL}{\mathcal{L}}

\newcommand{\mcT}{\mathcal{T}}

\newcommand{\mcV}{\mathcal{V}}

\newcommand{\mbE}{\mathbb{E}}

\newcommand{\bD}{\mathbf{D}}

\newcommand{\bQ}{\mathbf{Q}}

\newcommand{\bX}{\mathbf{X}}
\newcommand{\bY}{\mathbf{Y}}

\newcommand{\hF}{\hat{F}}

\newcommand{\hH}{\hat{H}}

\newcommand{\hV}{\hat{V}}
\newcommand{\hW}{\hat{W}}

\newcommand{\hp}{\hat{p}}

\newcommand{\hmu}{\hat{\mu}}

\newcommand{\tmuR}{\tilde{\mu}_{r_C}}

\newcommand{\bx}{\mathbf{x}}
\newcommand{\by}{\mathbf{y}}
\newcommand{\bz}{\mathbf{z}}

\newcommand{\bq}{\mathbf{q}}
\newcommand{\bk}{\mathbf{k}}

\newcommand{\bd}{\mathbf{d}}

\newcommand{\br}{\mathbf{r}}

\newcommand{\bn}{\mathbf{n}}

\newcommand{\hbF}{\hat{\mathbf{F}}}

\newcommand{\ba}{\mathbf{a}}
\newcommand{\bb}{\mathbf{b}}

\newcommand{\hbp}{\hat{\mathbf{p}}}
\newcommand{\hbq}{\hat{\mathbf{q}}}

\newcommand{\hbP}{\hat{\mathbf{P}}}
\newcommand{\hbQ}{\hat{\mathbf{Q}}}

%% file: table.tex
\begin{table*}[t]
\centering
\begin{tblr}{
	width = \textwidth,
	colspec = {|Q[c,m]|Q[c,m]|Q[c,m]|Q[c,m]|},
	vlines,
	hlines,
	cells   = {halign=c, valign=m},
	row{1}  = {bg=gray!10},
	column{1} = {bg=gray!10},
    colsep=3pt, rowsep=2pt
}
& GRW & CSL & DP \\

$\displaystyle \mcL_C [\rho_t]$ &
$\displaystyle -\sum_j \lambda_j \prt{\rho_t-T_j[\rho_t]}$ &
$\displaystyle -\frac{\gamma_{\rm CSL}}{2 m_0^2}\int \dd[3]{\bx} \comm{\hmu_{r_C}(\bx)}{\comm{\hmu_{r_C}(\bx)}{\rho_t}}$ &
$\displaystyle - \frac{G}{4\hbar} \int \dd[3]{\bx}\dd[3]{\by}\frac{\comm{\hmu_{r_C}(\bx)}{\comm{\hmu_{r_C}(\by)}{\rho_t}}}{\abs{\bx-\by}}$ \\

$\displaystyle  \dot{E}_t$ &
$\displaystyle \prt{\sum_j \frac{\lambda_j}{m_j}}\hbar^2 I[\sqrt{g_{r_C}}]$ &
$\displaystyle M m_0^{-2} \gamma_{\rm CSL} \hbar^2 I[g_{r_C}]$ &
$\displaystyle M \hbar G I_{\rm DP} [g_{r_C}]$ \\

$\displaystyle g_{r_C} (\bx)$ &
$\displaystyle \frac{1}{(2\pi r_C^2)^{3/2}}\exp{-\bx^2/(2 r_C^2)}$ &
$\displaystyle \frac{105}{32 \pi (3 r_C)^7}\prtq{9 r_C^2-\bx^2}_+^2$ &
$\displaystyle \frac{15}{8 \pi (\sqrt{7}r_C)^5}\prtq{7 r_C^2-\bx^2}_+$ \\

Gaussian increase &
0 &
47\% &
22\% \\
\end{tblr}

\caption{The first row of this table shows the specific form of $\mcL_C [\rho_t]$ for the various models, where $T_j [\rho_t] := \int \dd[3]{\bx}\sqrt{g_{r_C} (\bx-\hbq_j)}\rho_t\sqrt{g_{r_C} (\bx-\hbq_j)}$ and $\hmu_{r_C} (\bx) = \sum_j m_j g_{r_C} (\bx-\hbq_j)$.
The second row shows the explicit formula for the heating rate, where $m_0$ is a reference mass usually taken to be the proton mass, $G$ is Newton's constant, the $\lambda_j$ and $\gamma_{\rm CSL}$ are constants specific to the GRW~\cite[see pag. 305]{Review_Bassi2003Dynamical} and CSL~\cite[see pag. 491]{Review_Bassi2013Models} models, respectively, and $g_{r_C} (\bx)$ is the smearing distribution used to avoid the heating rate divergence, with the subscript $r_C$ denoting the smearing length defined as the square root of the distribution's variance. We also introduced a shorthand notation for the functionals $I[f]:=(1/2)\int\dd[3]{\bx} \abs{\nabla f(\bx)}^2$ and $I_{\rm DP}[f]:= (1/4)\int \dd[3]{\bx}\dd[3]{\by} \abs{\bx-\by}^{-1}[\nabla f(\bx)]\cdot[\nabla f(\by)]$.
The third row shows the distribution $g_{r_C}$ that minimizes the heating rate for a given value of the smearing length $r_C$, with $[x]_+:=\max\prtg{x,0}$.
Finally, the fourth row shows how much the heating rate increases by using the Gaussian in place of the optimal distribution.}
\label{tab:my-table}
\end{table*}